\documentclass[12pt,aps,nofootinbib,preprintnumbers]{revtex4-2}
\usepackage{amssymb}
\usepackage{amsmath}
\usepackage[colorlinks=true, pdfstartview=FitV, linkcolor=blue, citecolor=blue, urlcolor=blue]{hyperref}

\newcommand{\bp}{{\boldsymbol p}}
\newcommand{\tF}{\tilde{F}}
\newcommand{\calR}{\mathcal{R}}
\newcommand{\zero}{ {(0)} }
\newcommand{\one}{ {(1)} }
\newcommand{\two}{ {(2)} } 
\newcommand{\rme}{{\mathrm{e}}}
\newcommand{\rmi}{{\mathrm{i}}}
\newcommand{\rmd}{{\mathrm{d}}}

\begin{document}

\title{Nonlinear chiral kinetic theory}
\author{Kazuya~Mameda}
\affiliation{Department of Physics, Tokyo University of Science, Tokyo 162-8601, Japan}
\affiliation{RIKEN iTHEMS, RIKEN, Wako 351-0198, Japan}
\preprint{RIKEN-iTHEMS-Report-23}

\begin{abstract}
From quantum field theory, we derive the chiral kinetic theory involving nonlinear quantum corrections coupled with spacetime-dependent electromagnetic fields and fluid velocity gradients.
An equilibrium Wigner function determined by the kinetic equation verifies the nondissipativeness of the charge induced by the magneto-vortical coupling.
We reveal that this nonlinear chiral kinetic theory is consistent with the one-loop Euler--Heisenberg effective theory, indicating an indirect evidence of the trace anomaly in the kinetic theory.
We also argue a potential issue on the regularization, and demonstrate the availability of the point-splitting regularization in the nonlinear chiral kinetic theory.
\end{abstract} 

\maketitle

\section{Introduction}
The chiral kinetic theory (CKT) is one of the prominent theoretical tools to describe transport phenomena of massless degrees of freedom.
In this framework, a lot of transport phenomena are displayed with the Berry monopole~\cite{Stephanov:2012ki,Son:2012wh,Chen:2012ca}, as in the electron transport theory~\cite{Xiao:2009rm}.
A significant advantage of the CKT is the versatile applicability not only to  heavy-ion collisions~\cite{Liu:2021uhn,Fu:2021pok}, Weyl semimetal~\cite{Gorbar:2016qfh,Gorbar:2016ygi} and neutrino physics~\cite{Yamamoto:2015gzz,Yamamoto:2020zrs,Yamamoto:2021hjs}, but also to the photonic transport~\cite{Hattori:2020gqh,Huang:2020kik,Yamamoto:2017uul,Mameda:2022ojk}.
The CKT has also inspired us to elucidate many aspects in relativistic quantum transport, such as the Lorentz covariance~\cite{Chen:2014cla,Chen:2015gta,Hidaka:2016yjf}, collisional effects~\cite{Hidaka:2016yjf,Yang:2020hri,Weickgenannt:2021cuo,Lin:2021mvw}, the mass corrections~\cite{Gao:2019znl,Weickgenannt:2019dks,Hattori:2019ahi}, the strong magnetic field limit~\cite{Hattori:2016lqx,Sheng:2017lfu,Lin:2019fqo,Lin:2021sjw}, the different derivations~\cite{Manuel:2013zaa,Manuel:2014dza,Carignano:2018gqt,Carignano:2019zsh,Mueller:2017arw,Mueller:2017lzw}, and gravitational contributions~\cite{Liu:2018xip,Liu:2020flb,Hayata:2020sqz,Gao:2020gcf} (see also Ref.~\cite{Hidaka:2022dmn} and reference therein).

In spite of various developments, the usual CKT includes only the linear quantum correction.
One limitation of this linear CKT is found in the transport phenomena induced by the nonlinear coupling of background fields.
A particular example belonging to this category is the charge density of chiral fermions under external magnetic field and vortical field.
Such an induced charge is originally discovered from the diagrammatic computation based on the linear response theory~\cite{Hattori:2016njk}, and the agreement is found from the Dirac theory of a rotating fermions (for instance, see Ref.~\cite{Ebihara:2016fwa}).
Importantly, this charge generation is believed to be originated from quantum anomaly, and thus to be nondissipative~\cite{Kharzeev:2011ds}.
Nevertheless, the nondissipativeness cannot be verified within thermal field theory, including the linear response theory.
Indeed, the equilibration under magnetic field and rotation is subtle, since the coexistence of these external fields generates the drift force playing a role of an effective electric field.
The kinetic theory based on the Wigner function~\cite{Elze:1986qd} would provide a field-theoretical manifestation of the nondissipativeness, and thus the anomalous nature.
In this direction, the off-equilibrium formulation of the kinetic theory is required, beyond the near-equilibrium studies~\cite{Yang:2020mtz,Lin:2021sjw}.

Another limitation of the linear CKT is uncovered in the trace anomaly of quantum electrodynamics (QED), which is also the nonlinear quantum effect in the kinetic theory.
While the chiral anomaly is well known as a consequence of the Berry curvature, it is unobvious how the trace anomaly is interpreted in the kinetic description.
An important clue to answer this question is the consistency of the kinetic theory and quantum field theory.
Particularly, the CKT and the Euler--Heisenberg effective theory~\cite{Heisenberg:1936nmg,PhysRev.82.664} should inherit the same QED properties, since both theories describe fermionic dynamics under background electromagnetic fields.
Such a consistency is also a guiding principle in developing the CKT with nonlinear quantum corrections.

In this paper, based on quantum field theory, we formulate the nonlinear CKT, i.e., the CKT involving the nonlinear quantum correction coupled with spacetime-dependent electromagnetic and fluid velocity fields.
For this purpose, we derive the off-equilibrium Wigner function~\cite{Elze:1986qd} in the collisionless limit as a simple attempt.
Although the equilibrium state is not completely determined in the collisionless case, the frame-independence of the Wigner function provides a strong constraint for the equilibrium~\cite{Hayata:2020sqz}.
From an equilibrium Wigner function found in this way, we show the nondissipativeness of the magneto-vortical transport found in Ref.~\cite{Hattori:2016njk}.
We also find that the nonlinear CKT yields transport phenomena consistent with the Euler--Heisenberg effective theory.
This consistency further elucidates the kinetic encoding of the charge renormalization and the QED $\beta$-function, which is an indirect evidence of the trace anomaly in the CKT.

As a striking difference from the linear CKT, the nonlinear CKT bears an inevitable ultraviolet divergence to be properly regularized.
In this paper, we pose a potential issue on this regularization;
the competent techniques, such as Pauli--Villars regularization and dimensional regularization, are incompatible with the CKT.
Instead, we implements the point-splitting regularization~\cite{PhysRev.128.2425} in the nonlinear CKT.
Despite the violation of the translational invariance, this scheme is not only compatible with the Wigner function, but also helpful in elucidating the consistency with the Euler--Heisenberg theory.

This paper is organized as follows.
In Sec.~\ref{sec:R_eom}, we derive the off-equilibrium Wigner function at $O(\hbar^2)$, except for the distribution function.
In Sec.~\ref{sec:equilibrium}, analyzing the frame-dependence of the nonlinear CKT, we identify an equilibrium Wigner function.
In Sec.~\ref{sec:integral}, we demonstrate the computational manner of the momentum integral in the CKT, including the implementation of the point-splitting regularization.
In Sec.~\ref{sec:JT}, we evaluate the $O(\hbar^2)$ contributions to the equilibrium charge current and energy-momentum tensor.
In Sec.~\ref{sec:EH}, we show the consistency of the nonlinear CKT and the Euler--Heisenberg theory.
Section~\ref{sec:summary} is devoted to the summary of this paper.
We set $e=1$ in this paper unless otherwise stated, and use the mostly negative Minkowski metric.

\section{Nonlinear chiral kinetic theory}\label{sec:R_eom}

\subsection{Transport equations}
Based on quantum field theory, the transport theory is constructed from the Dyson-Schwinger equation for the Green's function.
When we consider virtual gauge fields, the corresponding equation for Dirac propagators yields the collisional kinetic theory.
This is important for pursuing the dynamical evolution in practical systems.
Nevertheless, since our present interest is to formulate the kinetic theory with nonlinear quantum corrections, through this paper, we only focus on the collisionless limit.

We consider the Dirac theory of fermion fields $\psi$ and $\bar{\psi}$ coupled with an external electromagnetic field $A_\mu$.
The two-point correlation functions $S_{\alpha\beta}^<(x,y) := \langle \bar\psi_\beta(y) \psi_\alpha(x) \rangle$ and $S_{\alpha\beta}^>(x,y) := \langle \psi_\alpha(x) \bar\psi_\beta(y) \rangle$ obey
\begin{equation}
\label{eq:Seq}
D_{x,\mu} S^{<} (x,y) = S^{>} (x,y) \overleftarrow{D}_{x,\mu} = 0 
\end{equation}
with $D_\mu \psi(x) := (\partial_\mu + \rmi A_\mu/\hbar)\psi(x)$ and $\bar\psi(x) \overleftarrow{D}_\mu := \psi(x) (\overleftarrow{\partial}_\mu - \rmi A_\mu/\hbar)$.
Note that here we implicitly enclosed the Wilson line, which ensures the gauge covariance of $S^\gtrless$.
This is equivalent to define the gauge covariant translation operator as $\psi(x+y) := \rme^{y\cdot D} \psi(x)$.
Fourier-transforming Eq.~\eqref{eq:Seq}, we get the transport equation of the Wigner function
\begin{equation}
\label{eq:W}
 W^\gtrless(x,p) 
 := \int_y \rme^{-\rmi p\cdot y/\hbar} S^\gtrless (x-y/2,x+y/2)
\end{equation}
with $\int_y := \int \rmd^4 y$.
The original transport equation of $W^\gtrless(x,p)$ contains the full quantum effect, and can be expanded in terms of $\hbar$~\cite{Elze:1986qd}.
This expansion is the same as that in terms of the spacetime gradient since $\hbar$ always accompanies a spacetime derivative.
The first nonlinear terms of $O(\hbar^2)$ thus emerge together with the second power of background electromagnetic fields and vortical fields, and their derivatives.
In the following analysis, we discuss only the lesser part $W(x,p):= W^<(x,p)$, which describes the kinetic theory of fermions.

In four-dimensional spacetime, the Wigner function can be decomposed with the basis of the Clifford algebra as 
\begin{equation}
 \label{eq:Clifford}
 W = \mathcal F + i\gamma^5\mathcal P+\gamma^\mu \mathcal V_\mu + \gamma^5\gamma^\mu \mathcal A_\mu+\tfrac{1}{2}\sigma^{\mu\nu}{\mathcal S}_{\mu\nu} ,
\end{equation}
where $\mathcal F$, $\mathcal P$, $\mathcal V_\mu$ $\mathcal A_\mu$ and $\mathcal S_{\mu\nu}$ are some coefficient fields dependent on $x^\mu$ and $p_\mu$.
For the transport equation of chiral fermions, the right-handed projection of $W(x,p)$ is decoupled (and so is the left-handed one) from other channels.
We denote this by
\begin{equation}
\calR(x,p) := \frac{1}{2} \mathrm{tr}[\gamma^\mu P_\mathrm{R} W (x,p)]
\end{equation}
with $P_\mathrm{R} := \frac{1}{2}(1+\gamma^5)$ and the trace is for the spinor indices.
The equations of motion for $\calR^\mu$ are derived as follows:
\begin{eqnarray}
 \label{eq:R1}
 & (\Delta_\mu + \hbar^2 P_\mu)\calR^\mu = 0  , \\ 
 \label{eq:R2}
 & (p_\mu + \hbar^2 Q_\mu )\calR^\mu = 0  , \\
 \label{eq:R3}
 & \hbar\varepsilon_{\mu\nu\rho\sigma}\Delta^{\rho}\calR^{\sigma}
 + 4\Bigl[
 		p_{[\mu} 
		 + \hbar^2 Q_{[\mu}
 	\Bigr]
 	 \calR_{\nu]}
  = 0  .
\end{eqnarray}
Here we defined $X_{[\mu}Y_{\nu]} := \frac{1}{2}(X_\mu Y_\nu -X_\nu Y_\mu)$, the Levi-Civita tensor with $\varepsilon^{0123} = 1$ and the following differential operators:
\begin{equation}
 \Delta_\mu 
  = \partial_\mu -F_{\mu\lambda} \partial_p^\lambda    ,\quad
  P_\mu   
  = \frac{1}{24} (\partial_p\cdot\partial)^2 F_{\mu\nu} \partial_p^\nu   ,\quad
 Q_\mu 
  = - \frac{1}{12}\partial_p\cdot\partial F_{\mu\nu}\partial^\nu_p  .
\end{equation}
Contracting Eq.~\eqref{eq:R3} with $p^\nu$ and using Eq.~\eqref{eq:R2}, we get the useful equation
\begin{equation}
\label{eq:R4}
 p^2 \calR_\mu 
 = \frac{\hbar}{2} \varepsilon_{\mu\nu\rho\sigma}p^\nu \Delta^\rho \calR^\sigma
 + 2\hbar^2  p^\nu Q_{[\mu} \calR_{\nu]}
 - \hbar^2 p_\mu Q\cdot\calR  .
\end{equation}

Once $\calR^\mu$ is determined from the above equations of motion, we can compute physical quantities.
By implementing the inverse Wigner transformation of two point functions, the charge current, energy-momentum tensor and spin tensor are expressed with $\calR^\mu$, as follows:
\begin{eqnarray}
\label{eq:J_mu}
 J^\mu (x,y) &=& 2\int_p  \rme^{\rmi p\cdot y/\hbar} \calR^\mu(x,p) ,\\
\label{eq:T_munu}
 T^{\mu\nu}(x,y) &=& 2\int_p  \rme^{\rmi p\cdot y/\hbar} 
 \Bigl[
 	p^{(\mu} \calR^{\nu)}(x,p) + \hbar^2 Q^{(\mu} \calR^{\nu)}(x,p) 
 \Bigr] ,\\
\label{eq:S_munurho}
 S^{\mu\nu\rho}(x,y) &=& - 2\hbar\, \varepsilon^{\mu\nu\rho\sigma} \int_p  \rme^{\rmi p\cdot y/\hbar} \calR_\sigma(x,p)
\end{eqnarray}
with $\int_p := \int \frac{\rmd^4 p}{(2\pi)^4}$ and $X_{(\mu}Y_{\nu)} := \frac{1}{2}(X_\mu Y_\nu + X_\nu Y_\mu)$.
In Appendix~\ref{app:JTS}, we derive Eqs.~\eqref{eq:J_mu}-\eqref{eq:S_munurho} from the two-point functions.
In the usual analysis with the Wigner function approach, the above quantities are defined in the $y\to 0$ limit.
However, this parameter $y$ plays a role of the ultraviolet regulator when we implement the point-splitting regularization.
For this reason, hereafter we keep $y$ finite.

From these expressions~\eqref{eq:J_mu}-\eqref{eq:S_munurho}, it is manifested that  Eqs.~\eqref{eq:R1}-\eqref{eq:R3} correspond to the Ward identities which massless fermions should respect.
The first equation~\eqref{eq:R1} is related to charge conservation, and thus interpreted as the kinetic equation, which determines the distribution function in $\calR^\mu$.
The latter two~\eqref{eq:R2} and~\eqref{eq:R3} imply the conformal invariance and the Lorentz invariance (i.e., angular momentum conservation), respectively.
These two determine the off-equilibrium Wigner function, except for the distribution function.

\subsection{Solution up to $O(\hbar^2)$}\label{sec:solution}

In the following, we look for the solution of Eqs.~\eqref{eq:R2}-\eqref{eq:R3} and ~\eqref{eq:R4}, with the parametrization:
\begin{equation}
 \calR^\mu = \calR^\mu_\zero + \hbar\calR^\mu_\one+ \hbar^2 \calR_\two^\mu  .
\end{equation}
For the latter computation of the nonlinear solution $\calR^\mu_\two$, let us first briefly review the $O(\hbar^0)$ and $O(\hbar)$ parts~\cite{Hidaka:2016yjf}.
The $O(\hbar^0)$ solution is readily found from Eqs.~\eqref{eq:R2} and~\eqref{eq:R4} as
\begin{equation}
\label{eq:Rmu0}
 \calR^\mu_\zero = 2\pi\delta(p^2) p^\mu f_\zero  ,
\end{equation}
where $f_\zero$ is a function that satisfies $\delta(p^2) p^2 f_\zero = 0$.
The delta function $\delta(p^2)$ represents the on-shell condition of the chiral fermion: $p^2 = (p_0)^2 - |\bp|^2 = 0$.
This $f_\zero$ has both particle and antiparticle contributions.
At equilibrium, $f_\zero$ is the Fermi distribution function, with which the Wigner function $\calR^\mu_\zero$ reproduces the usual lesser propagator~\cite{Bellac:2011kqa}.

Let us solve the first-order part.
Inserting the zeroth-order solution~\eqref{eq:Rmu0} into Eq.~\eqref{eq:R4}, we get the first-order correction as
\begin{equation}
 \calR^\mu_\one 
 = 2\pi\delta(p^2) 
 	\biggl[
 		\widetilde{\calR}^\mu_\one 
 		- \frac{1}{p^2}\tF^{\mu\nu}  p_\nu f_\zero
 	\biggr] 
\end{equation}
with $\tF_{\mu\nu}=\frac{1}{2}\varepsilon_{\mu\nu\rho\sigma} F^{\rho\sigma}$.
The second term is apparently singular, but it accounts for the chiral anomaly in the CKT.
Also, we emphasize the existence of the first term, which is admitted as long as it satisfies $\delta(p^2) p^2 \widetilde{\calR}^\mu_\one = 0$ and $\delta(p^2) p\cdot\widetilde{\calR}_\one = 0$.
This extra term is determined from Eq.~\eqref{eq:R3} at $O(\hbar)$, as follows:
\begin{equation}
 \begin{split}
  \widetilde{\calR}_\mu^\one\delta(p^2)
  = \delta(p^2)
 	\biggl[
 		p_\mu \frac{n\cdot \widetilde{\calR}_\one}{p\cdot n} 
 		+ \frac{\varepsilon_{\mu\nu\rho\sigma}p^\rho n^\sigma}{2p\cdot n} 
 			\Delta^\nu f_\zero
 	\biggr]  ,
 \end{split}
\end{equation}
where we introduce an arbitrary vector field $n^\mu(x)$.
Thus, the first correction part is given by
\begin{equation}
\label{eq:Rmu1}
 \calR^\mu_\one 
 = 2\pi\delta(p^2)
 \biggl[
  p^\mu f_\one 
  + 
 	\biggl(
 		\Sigma_n^{\mu\nu}\Delta_\nu
  		- \frac{1}{p^2}\tF^{\mu\nu} p_\nu 
  	\biggr) f_\zero
 \biggr]  ,
\end{equation}
where we define
\begin{equation}
\label{eq:f1_Sigman}
  f_\one
  := \frac{n\cdot\widetilde{\calR}^\one}{p\cdot n}  ,
   \quad 
  \Sigma_n^{\mu\nu} 
   := \frac{\varepsilon^{\mu\nu\rho\sigma} p_\rho n_\sigma}{2p\cdot n}  .
\end{equation}
This tensor $\Sigma_n^{\mu\nu}$ corresponds to the spin of chiral fermions and $n^\mu$ is the degrees of freedom for the frame choice of the spin~\cite{Chen:2014cla,Chen:2015gta}.
It is worth mentioning that $\delta(p^2)p^2\widetilde{\calR}^\mu_\one = 0$ implies $\delta(p^2)p^2 f_\one = 0$.
Such a nonsingular condition for $f_\one$ is important, in particular, when we determine the equilibrium form of $f_\one$.
Also, $\delta(p^2)p^2 f_\one = 0$ ensures that the above solution~\eqref{eq:Rmu1} fulfills Eqs.~\eqref{eq:R2} and~\eqref{eq:R4}.

In a totally parallel manner, we can solve the second-order part $\calR^\mu_\two$. The derivation is shown in Appendix~\ref{app:2nd_solution} (see also Ref.~\cite{Hayata:2020sqz}).
The result is
\begin{equation}
 \begin{split}
 \label{eq:Rmu2}
  \calR_\mu^\two
	 & =  2\pi\delta(p^2)
 	  	\biggl[
 	  		p_\mu f_\two
 	  		+ \biggl(
 	  			\Sigma_{\mu\nu}^u \Delta^\nu 
 	  			- \frac{1}{p^2}\tF_{\mu\nu}p^\nu
 	  		 \biggr) f_\one
 	  		- \Sigma_{\mu\nu}^u\varepsilon^{\nu\rho\sigma\lambda}
				\Delta_{\rho} \frac{n_\sigma}{2p\cdot n}\Delta_\lambda f_\zero
	    \biggr] \\
	  &\quad
	     + \frac{2\pi}{p^2}
	     	\biggl[
	     		- p_\mu Q\cdot p 
	     		+ 2 p^\nu Q_{[\mu} p_{\nu]} 
	     	\biggr]
	     		f_\zero \delta(p^2)\\
	  &\quad
	     	+ 2\pi\frac{\delta(p^2)}{p^2}
	     	\biggl(
	     	 \frac{1}{2}\varepsilon_{\mu\nu\rho\sigma} p^\nu \Delta^\rho
	  		  + \frac{p_\mu p^\nu}{p^2} \tF_{\nu\sigma}
	  		  - \tF_{\mu\sigma}
	  		\biggr)
  			\biggl(
  				\Sigma^{\sigma\lambda}_n \Delta_\lambda 
  				- \frac{1}{p^2}\tF^{\sigma\lambda}  p_\lambda 
  			\biggr)f_\zero \\
  	 &\quad
 	+ 2\pi\frac{\delta(p^2)}{p^2}\Sigma_{\mu\nu}^u 
		\biggl[
			\Delta_\alpha\Sigma^{\alpha\nu}_n
			+\frac{n_\alpha}{p\cdot n}\tF^{\alpha\nu}
			+\frac{1}{p^2}\tF^{\nu\lambda}p_\lambda
		\biggr]p\cdot\Delta f_\zero  ,
 \end{split}
\end{equation}
where $\Delta_\mu$ and $Q_\mu$ operate all on the right.
Here, another vector field $u^\mu$ and spin tensor $\Sigma^{\mu\nu}_u$ are introduced, similarly to $n^\mu$ and $\Sigma^{\mu\nu}_n$ in $\calR^\mu_\one$.
The new factor $f_\two$ is the second-order counterpart of $f_\one$, and is required to satisfy the nonsingular condition $\delta(p^2)p^2 f_\two = 0$.
For $u^\mu=n^\mu$, the above solution $\calR^\mu = \calR^\mu_\zero + \hbar\calR^\mu_\one + \hbar^2\calR^\mu_\two$ can be recast in a simpler form.
Then, $f_{\zero}$, $f_{\one}$ and $f_{\two}$ in $\calR^\mu$ are totally combined as the single function $f:=f_\zero + \hbar f_\one + \hbar^2 f_\two$, as are so in the gravitational case~\cite{Hayata:2020sqz}. 
Inserting this $\calR^\mu$ into Eq.~\eqref{eq:R1}, we get the $n^\mu$-dependent nonlinear chiral kinetic equation to determine the single distribution function $f$.
Such a structure is the same as the linear chiral kinetic equation.
For this reason, $u^\mu$ could be regarded as the degrees of freedom for the Lorentz transformation.
On the other hand, the above interpretation of $u^\mu$ is inapplicable for $u^\mu\neq n^\mu$, and thus the physical meaning of $u^\mu$ is not completely identified.
To address this problem, we should study the Lorentz transformation up to $O(\hbar^2)$ in quantum field theory~\cite{Hidaka:2016yjf}.
Although this is an important task to manifest the nonlinear-order side-jump effect~\cite{Chen:2014cla,Chen:2015gta}, we will analyze it in a future publication.
Hereafter, we call both $n^\mu$ and $u^\mu$ the frame vectors.

\section{Equilibrium}\label{sec:equilibrium}

\subsection{Frame-dependence}\label{sec:frame}
As is well known, the CKT depends on the frame vectors $n^\mu$ and $u^\mu$.
Since the frames are auxiliary fields to obtain the solutions~\eqref{eq:Rmu1} and~\eqref{eq:Rmu2}, however, physical quantities should be independent of the frames, and so is $\calR^\mu$.
On the other hand, the distribution function depends on the frame~\cite{Chen:2015gta}.
In the linear CKT, the frame transformation law of $f_\one$ is determined by imposing $\calR^\mu_\one$ keeps frame-independent~\cite{Hidaka:2016yjf}.
Similarly, in the nonlinear CKT, we can compute the transformation law of $f_\two$ from the frame-independence of $\calR^\mu_\two$~\cite{Hayata:2020sqz}.
Let us first focus on the variation in terms of $n^\mu$.
Suppose that we take the transformation of the frame vector as $n^\mu \to n'^\mu$.
Then the corresponding transformation of the distribution function is written as $f_\one \to f_\one + \delta_n f_\one$, $f_\two \to f_\two + \delta_n f_\two$.
It is worthwhile to mention that the variations $\delta_n f_{\one,\two}$ should be nonsingular because so are $f_{\one,\two}$.
That is, we impose $\delta(p^2) p^2 \delta_n f_\one = \delta(p^2) p^2 \delta_n f_\two = 0$.

The frame-independence of $\calR_\one^\mu$ is represented as $\calR_\one^\mu|_{n'} - \calR_\one^\mu|_{n} = 0$, where $\calR_\one^\mu|_{n}$ is the Wigner function in Eq.~\eqref{eq:Rmu1} with a frame $n^\mu$.
From this equation, we determine the transformation law of $f_\one$, as follows:~\cite{Chen:2015gta,Hidaka:2016yjf}
\begin{equation}
 \begin{split}
 \label{eq:delta_f1}
  \delta_n f_\one 
   & = - \frac{n^\mu}{p\cdot n}\Sigma_{\mu\nu}^{n'} \Delta^\nu f_\zero 
   		+ p^2 \delta_n g_\one   ,
 \end{split}
\end{equation}
where $\delta_n g_\one$ is a nonsingular scalar fulfills $\delta(p^2) p^2 \delta_n g_\one = 0$.
In the linear CKT, this $\delta_n g_\one$ can be ignored;
such a term does not affect $\calR^\mu_\one$.
This is, however, not the case in the nonlinear CKT.
Indeed, from a similar but more complicated evaluation for $\calR^\mu_\two$, we obtain the variation of $f_\two$ as
\begin{equation}
 \begin{split}
 \label{eq:deltan_f2}
  \delta_n f_\two
	& =   \Sigma_{\mu\nu}^u 
		\biggl[
			\Delta^\mu \frac{\varepsilon^{\nu\rho\alpha\beta}  n_\alpha n'_\beta}{2p\cdot n\,p\cdot n'}
   			\Delta_\rho f_\zero 
   			- F^{\mu\nu} \delta_n g_\one
   		\biggr] \\
   	&\quad
		+\frac{1}{p^2}
   		\biggl[
   			\Sigma^u_{\mu\nu} \Delta^\mu 
   			- \tF_{\mu\nu}
   		\biggl(
   			\frac{p^\mu}{p^2}
   			-\frac{u^\mu}{p\cdot u}
   		\biggr)
   		\biggr]
   			\Sigma^{\nu\lambda}_{n'}\frac{n_\lambda}{p\cdot n}p\cdot\Delta f_\zero
   		 .
 \end{split}
\end{equation}
which involves $\delta_n g_\one$.
The same analysis can be performed for the variation in terms of $u^\mu$.
Then, we find $\delta_u f_\one = 0$ and
\begin{equation}
\label{eq:deltau_f2}
 \delta_u f_\two 
 = -\frac{u^\mu}{p\cdot u} \Sigma^{u'}_{\mu\nu}
 	\biggl[
 		  \Delta^\nu f_\one
	- \varepsilon^{\nu\rho\sigma\lambda}
 	  			\Delta_\rho \frac{n_\sigma}{2p\cdot n} \Delta_\lambda f_\zero
 	 + \frac{1}{p^2}
 	 \biggl(
 	  			\Delta_\alpha\Sigma_n^{\alpha\nu}
 	  			+\frac{n_\alpha}{p\cdot n}\tF^{\alpha\nu}
 	  			+\frac{1}{p^2}\tF^{\nu\lambda}p_\lambda
 	  \biggr)
 	  	p\cdot\Delta f_\zero
   	\biggr]  .
\end{equation}

\subsection{Equilibrium Wigner function}\label{sec:Weq}
Let us apply the above argument to the equilibrium solution of the nonlinear CKT.
In the collisionless case, the kinetic theory itself cannot generally determine equilibrium.
The frame transformation laws~\eqref{eq:delta_f1}-\eqref{eq:deltau_f2} however provide strong constraints to fix the equilibrium distribution functions.
To illustrate this fact, let us here employ the equilibrium distribution function so that the classical Wigner function~\eqref{eq:Rmu0} is reproduced as the well-known form of the lesser Green's function of free fermions, that is,
\begin{eqnarray}
\label{eq:equilibrium_f0}
 & f_\zero = \epsilon(p_0) \, n_F(-\mu + p\cdot\xi)  ,\quad
\partial_\mu \alpha - F_{\mu\nu}\beta^\nu = 0  , \quad
 \partial_\mu \beta_\nu + \partial_\nu \beta_\mu = 0  ,
\end{eqnarray}
where we define $\epsilon(x) := \theta(x) - \theta(-x)$ with the step function $\theta(x)$, and the Fermi distribution function $n_F(x) := (\rme^{\beta x}+1)^{-1}$.
The parameters $\alpha$ and $\beta^\mu$ are defined as $\alpha=-\beta\mu$, $\beta^\mu=\beta\xi^\mu$ and $\xi\cdot\xi = 1$ with chemical potential $\mu$, inverse temperature $\beta$ and fluid velocity $\xi^\mu$.
The Wigner function $\calR^\mu_\zero$ with this $f_\zero$ in fact solves the classical kinetic equation~\eqref{eq:R1}:
$\Delta\cdot\calR_\zero = 2\pi\delta(p^2) f'_\zero p^\mu (\partial_\mu \alpha + p^\nu\partial_\mu \beta_\nu - F_{\mu\nu} \beta^\nu) = 0$ with $f'_\zero = \rmd f_\zero(x)/\rmd x$ and $x=\alpha+\beta\cdot p$.

Then, using the above $f_\zero$, we compute the transformation law of $f_\one$ and $f_\two$.
From Eq.~\eqref{eq:delta_f1}, we obtain
\begin{equation}
 \begin{split}
  \delta_n f_\one 
  & = f'_\zero \frac{1}{2}(\Sigma^{\nu\rho}_{n'}-\Sigma^{\nu\rho}_{n})
  			\partial_\nu\beta_\rho
  	 + p^2
  	 \biggl[
  	 	\delta_n g_\one 
  	 	- f_\zero' \frac{\varepsilon^{\mu\nu\alpha\beta}n_\alpha n_\beta'}{4p\cdot n p\cdot n'} \partial_\nu\beta_\rho
  	 \biggr]  .
 \end{split}
\end{equation}
The above equation holds when we choose
\begin{equation}
 \label{eq:f1}
 f_{\one} 
 = f_\zero'\frac{1}{2}\Sigma_n^{\mu\nu}\partial_\mu\beta_\nu  ,\quad
 \delta_n g_\one 
 = f_\zero'\frac{\varepsilon^{\mu\nu\alpha\beta}n_\alpha n_\beta'}{4p\cdot n p\cdot n'} \partial_\nu\beta_\rho  .
\end{equation}
Similarly, the variations of $f_\two$ are calculated as follows:
\begin{equation}
\begin{split}
\delta_n f_\two
    & = \frac{1}{4}\Sigma_{\mu\nu}^u \Delta^\mu 
		\varepsilon^{\nu\beta\rho\lambda}\,\biggl(\frac{n'_\beta}{p\cdot n'}-\frac{n_\beta}{p\cdot n}\biggr)
		\partial_\rho\beta_\lambda f'_\zero  ,\\
\delta_u f_\two
	&= \frac{1}{4}(\Sigma_{\mu\nu}^{u'}-\Sigma_{\mu\nu}^u) \Delta^\mu 
		\varepsilon^{\nu\beta\rho\lambda}
		\frac{n_\beta}{p\cdot n}
		\partial_\rho\beta_\lambda f'_\zero  .
\end{split}
\end{equation}
We note that all singular terms with $(p^2)^{-1}$ or $(p^2)^{-2}$ in Eqs.~\eqref{eq:deltan_f2} and~\eqref{eq:deltau_f2} disappear, thanks to $p\cdot\Delta f_\zero =0$.
The above equations indicate that the second-order quantum correction $f_\two$ may be deduced as
\begin{equation}
 \label{eq:f2}
 f_\two
 = \Sigma^u_{\mu\nu}  \Delta^\mu 
 	\biggl(
 		f'_\zero
 		\frac{\varepsilon^{\nu\rho\sigma\lambda}}{4\,p\cdot n}
 		n_\rho \partial_\sigma\beta_\lambda 
 	\biggr) 
 	+ \phi_\two  .
\end{equation}
Here $\phi_\two$ is a frame-independent term in the equilibrium distribution function.
Such an ambiguity in $f_\two$ cannot be determined in the present framework, which ignore the collisional effect.

At the equilibrium we found above, the Wigner function~\eqref{eq:Rmu2} is reduced.
First, we assume $\phi_\two =0$ for simplicity.
Plugging Eqs.~\eqref{eq:f1} and~\eqref{eq:f2} into~Eq.\eqref{eq:Rmu2}, one can show that the frame-dependence of $\calR^\mu_\two$ is totally compensated, as it should.
Eventually, Eq.~\eqref{eq:Rmu2} is recast into the four different pieces as $ \calR^\two_\mu = \calR^{(\partial F)}_\mu+\calR^{(FF)}_\mu + \calR^{(F\omega)}_\mu + \calR^{(\omega\omega)}_\mu$ with
\begin{align}
\label{eq:R_delF}
 \calR^{(\partial F)}_\mu
& = 2\pi\frac{\delta(p^2)}{p^2}\cdot \frac{1}{12}
 	\Biggl[
 		p_\mu f_\zero \biggl(-\frac{8}{p^2}\biggr)
 			 p^\rho\partial^\lambda F_{\rho\lambda} 
 		+p_\mu f_\zero'
 		\biggl(
 			\partial^\rho F_{\rho\lambda} \beta^\lambda
 			-\frac{4}{p^2}
 					p^\rho p\cdot\partial F_{\rho\lambda}\beta^\lambda
 		\biggr) \nonumber\\
 & \qquad\qquad\quad
 		+p_\mu f_\zero''
 		\biggl(
 			2 p^\rho\beta\cdot\partial F_{\rho\lambda}\beta^\lambda
 		\biggr) 
	+ f_\zero
		\biggl(
			8 \partial^\lambda F_{\mu\lambda}
			-\frac{8}{p^2} p\cdot\partial F_{\mu\lambda}p^\lambda
		\biggr)\nonumber\\
 & \qquad\qquad\quad
	+ f'_\zero
		\biggl(
			p\cdot\partial F_{\mu\lambda}\beta^\lambda
			+p^\nu \partial_\mu F_{\nu\lambda}\beta^\lambda
		\biggr) 
 	+ f_\zero''
		\biggl(
			-p^2 \beta\cdot\partial F_{\mu\lambda} \beta^\lambda
		\biggr) 		
 	\Biggr] \\
\label{eq:R_FF}
 \calR^{(FF)}_\mu
 &= 2\pi\frac{\delta(p^2)}{(p^2)^2}\,
 		\cdot 2
		\biggl(
			-\frac{p_\mu p^\nu}{p^2} F_{\nu\sigma} 
			+ F_{\mu\sigma}
		\biggr) F^{\sigma\lambda}p_\lambda f_\zero  ,\\
\label{eq:R_Fo}
 \calR^{(F\omega)}_\mu
 &= 2\pi\frac{\delta(p^2)}{p^2}
 			\biggl(
				-p_\mu\frac{ p^\nu p_\rho}{p^2} \omega_{\nu\sigma} \tF^{\sigma\rho}
				+ \frac{3}{4} \omega_{\mu\sigma} \tF^{\sigma\nu} p_\nu
				+ \frac{1}{4} \tF_{\mu\sigma} \omega^{\sigma\nu} p_\nu
			\biggr)  f_\zero  ,\\
\label{eq:R_oo}
 \calR^{(\omega\omega)}_\mu
 &=2\pi\delta(p^2) \cdot \frac{1}{4}
	 	\biggl(			
			 p_\mu\frac{p^\nu p^\rho}{p^2}\omega_{\nu\sigma}{\omega_\rho}^\sigma
			- \omega_{\mu\sigma} {\omega_\nu}^\sigma p^\nu
		\biggr) f_\zero''  ,
\end{align}
where we introduce $\omega^{\mu\nu} := \frac{\beta^{-1}}{2}\varepsilon^{\mu\nu\rho\sigma}\partial_\rho\beta_\sigma$.
We also note that the derivative of vorticity disappears, i.e., $\calR^\mu_{(\partial \omega)}=0$, owing to the identity $\partial_\mu \partial_{\nu}\beta_{\rho} = 0$ for the Killing vector $\beta_\rho$.

At this point, it is not guaranteed that the above $\calR^\mu$ is really an equilibrium Wigner function, because we have not yet analyzed the $O(\hbar^2)$ part of the kinetic equation~\eqref{eq:R1}%
~\footnote{%
One can readily check that the $O(\hbar)$ part of Eq.~\eqref{eq:R1} holds for the linear-order solution~\eqref{eq:Rmu1}.
}.
Plugging Eqs.~\eqref{eq:Rmu0} and~\eqref{eq:R_delF}-\eqref{eq:R_oo} to the kinetic equation~\eqref{eq:R1} and carrying out a tedious computation, we arrive at
\begin{equation}
\begin{split}
\label{eq:keq}
 \delta(p^2)
  	\biggl[
		\biggl(
			\frac{f''_\zero}{6p^2} \partial^\mu \beta^\rho  p^\nu p_\rho 
			-\frac{f''_\zero}{8} \partial^\mu \beta^\nu 
 			-\frac{f_\zero'''}{12}  \partial^\mu \beta^\rho \beta^\nu p_\rho
 		\biggr)
 			 \beta\cdot\partial 
 		+\frac{f_\zero''}{24} p^\mu \beta^\nu \beta^\rho \beta^\sigma \partial_\rho\partial_\sigma 
 	\biggr]
 		F_{\mu\nu}
 	= 0 .
\end{split}
\end{equation}
Using $\beta^\rho\beta^\sigma \partial_\rho \partial_\sigma F_{\mu\nu} = \beta\cdot\partial(\beta\cdot\partial F_{\mu\nu}) - (\beta\cdot\partial\beta_\sigma) \partial^\sigma F_{\mu\nu}$, we find that all the terms in the above kinetic equation contain $\beta\cdot\partial F_{\mu\nu}$ or $\beta\cdot\partial \beta_\mu$.
As long as we consider a finite $F_{\mu\nu}$, hence, the above reduced kinetic equation implies that either of the following conditions should be fulfilled:%
~\footnote{%
Note that $\partial_\mu \beta_\nu = 0$ is an equilibrium condition.
In this case, $\beta\cdot \partial F_{\mu\nu} = 0$ automatically holds because of $0=\partial_{[\mu}\partial_{\nu]}\alpha = \partial_{[\mu}(F_{\nu]\lambda}\beta^\lambda)$.
This condition is however a special case of the condition~\eqref{eq:equilibrium1}.
}
\begin{subequations}
\label{eq:equilibrium}
\begin{align}
\label{eq:equilibrium1}
 & 1)\quad \beta\cdot\partial F_{\mu\nu} = 0  ,\quad
 \beta\cdot\partial \beta_\mu = 0  ,\\
\label{eq:equilibrium2}
 & 2)\quad \partial_\lambda F_{\mu\nu} = 0  .
\end{align}
\end{subequations}
These are the additional equilibrium conditions on top of those in Eq.~\eqref{eq:equilibrium_f0}.
The meaning of the condition~\eqref{eq:equilibrium1} is understandable when we take $\xi^\mu = (1,\boldsymbol{0})$.
The first equation in Eq.~\eqref{eq:equilibrium1} implies the time-independence of background electromagnetic fields.
The second means that the background fluid has no acceleration, or equivalently, there is no temperature gradient: $0= \beta\cdot\partial\beta_\mu = -\beta \partial_\mu \beta$ with $\beta :=\sqrt{\beta\cdot\beta}$.
On the other hand, the acceleration term is admitted under the condition~\eqref{eq:equilibrium2}, where electromagnetic fields are constant.
This is the case employed in Ref.~\cite{Yang:2020mtz}.

We here discuss the case with $\phi_\two \neq 0$ in Eq.~\eqref{eq:f2}.
One can readily check that in this case the extra term $\delta(p^2)p\cdot\Delta \phi_\two$ emerges in the kinetic equation~\eqref{eq:keq}.
However, the singular term with $p^{-2}$ cannot be eliminated by the $\phi_\two$ term, since $\delta(p^2) \phi_\two = 0$ is required from the nonsingular condition $\delta(p^2) f_\two = 0$.
Moreover, the other terms in Eq.~\eqref{eq:keq} are not canceled by the $\phi_\two$ term.
Hence, $\delta(p^2) p\cdot\Delta \phi_\two=0$ is demanded.
As the simplest choice, we may take $\phi_\two = 0$ hereafter.
This is a difference from the CKT in curved spacetime;
under a weak static gravitational field, a finite $\phi_\two$ is required for the realization of an equilibrium~\cite{Hayata:2020sqz}.

\section{Momentum integral}\label{sec:integral}

\subsection{Regularization}\label{sec:reg}
The equilibrium physical quantities are computed as the momentum integral with the Winger function in Eqs.~\eqref{eq:R_delF}-\eqref{eq:R_oo} with the distribution function~\eqref{eq:equilibrium_f0} under the condition~\eqref{eq:equilibrium}.
Before the computation, we demonstrate how to evaluate the momentum integrals.
The integrals that we encounter in the following section are generally written as
\begin{equation}
\label{eq:integral_general}
 \int_p 2\pi\frac{\rmd^l \delta(p^2)}{(\rmd p^2)^l} p^{\mu_1}\cdots p^{\mu_j} \frac{\rmd^k f_\zero (p_0)}{\rmd p_0^k} \rme^{\rmi p\cdot y/\hbar}
\end{equation}
with $f_\zero$ given by Eq.~\eqref{eq:equilibrium_f0}.
Here we replaced the singular factor $(p^2)^{-l}$ in the Wigner functions with the derivative of $\delta(p^2)$, through the identity $l!\delta(x) = (-x)^l\rmd^l\delta(x)/\rmd x^l$.

For the latter convenience, we here decompose Eq.~\eqref{eq:equilibrium_f0} into the vacuum and matter parts as $f_\zero(p_0) = f_{\zero\mathrm{vac}}(p_0) +  f_{\zero\mathrm{mat}}(p_0)$ with $f_{\zero\mathrm{vac}}(p_0):= -\theta(-p_0)$ and  $f_{\zero\mathrm{mat}}(p_0):=\theta(p_0)n_F(p_0-\mu) + \theta(-p_0)n_F(-p_0+\mu)$.
In Eq.~\eqref{eq:integral_general} the former may result in the divergence at the ultraviolet regime $p_0\sim -\infty$ unless $k\geq 1$.
For this divergence, the parameter $y^\mu$ plays a role of the cutoff scale.
This is nothing but the point-splitting regularization.
On the other hand, the latter does not require such a regulation.
Therefore, in the following, we evaluate these two contributions in different ways;
for the vacuum contributions, we keep $y$ finite so that the point-splitting regularization is implemented, but for the matter part we take $y\to 0$ before integration%
~\footnote{%
The point-splitting regularization with $n_F(p_0\mp\mu)$ would in principle be possible, but is not so easy as that of the vacuum;
due to the pole at $p_0 = \pm \mu + \rmi (2n+1)\pi T$ for $n=0,\pm 1,\cdots$, it is nontrivial to perform the Wick rotation, which is required in implementing the point-splitting regularization.%
}.
It should also be emphasized that we face no infrared divergence in Eq.~\eqref{eq:integral_general}, thanks to the cancellation of those from the vacuum and matter parts.

We comment on the regularization in the CKT.
In usual quantum field theory, when we regularize a divergent integral, it is preferred to choose a regularization scheme to respect the gauge, Lorentz, and translational invariances.
It is, however, not so easy to find out such an appropriate scheme for Eq.~\eqref{eq:integral_general}.
For instance, the Pauli--Villars scheme is obviously unsuitable, since the CKT possesses no mass parameter;
a Pauli--Villars regulator would be useful for the kinetic theory of massive fermions~\cite{Gao:2019znl,Weickgenannt:2019dks,Hattori:2019ahi}.
Dimensional regularization is also incompatible with the CKT, since $\varepsilon^{\mu\nu\rho\sigma}$ and $\gamma^5$ cannot be extended straightforwardly in a general $d$-dimensional spacetime~\cite{tHooft:1972tcz}.
Indeed, the Wigner functions derived in Secs.~\ref{sec:R_eom}-\ref{sec:equilibrium} are no longer correct in $d\neq 4$ dimensions, for the following two reasons.
First, the Clifford basis decomposition~\eqref{eq:Clifford} is unjustified in $d\neq 4$ dimensions.
This implies that our starting point at Eqs.~\eqref{eq:R1}-\eqref{eq:R3} is modified.
Second, the uselessness of the Schouten identity in $d\neq 4$ dimensions brings a lot of extra singular terms with $p^{-2}$ in intermediate steps of calculation.
Then we would not derive the solution that satisfies appropriate conditions, such as $\delta(p^2)p^2 f_\two = 0$.

The above circumstance compels us to choose a regularization scheme that sacrifices at least one symmetry.
Among such schemes, the point-splitting regularization is compatible with the Wigner function because the point-splitting parameter is naturally introduced as $y^\mu$, as shown in the charge current~\eqref{eq:J_mu} and the energy-momentum tensor~\eqref{eq:T_munu}.
This is the reason why we employ the point-splitting regularization in this paper.
Although this scheme in general violates the translational invariance (namely, $\partial_\mu T^{\mu\nu} + F^{\mu\nu}J_\mu \neq 0$), it can reveal the consistency with the Euler--Heisenberg theory, as we discuss later.
The analysis with a more appropriate regularization will be shown in feature publication.

\subsection{Matter part}
We demonstrate how to compute the matter part in Eq.~\eqref{eq:integral_general}.
We perform first the integral in terms of $p_0$ and then $p_i$.
In this way, by decomposing each $p^\mu$ into the transverse component to $\xi^\mu:=(1,\boldsymbol{0})$ and the longitudinal one, we can replace integrands with nonvanishing tensor form;
for instance, $p_{\alpha}\to p_{0}\xi_{\alpha}$ and $p_{\alpha}p_{\beta} \to (p_{0})^{2}\xi_{\alpha}\xi_{\beta} + \frac{\bp^{2}}{3}\Delta_{\alpha\beta}$ with the transverse projector $\Delta^{\mu\nu}:=\xi^{\mu}\xi^{\nu}-g^{\mu\nu}$.
Performing the tensor decomposition of the integrands, we express Eq.~\eqref{eq:integral_general} as the linear combination of
\begin{equation}
\label{eq:Il_nmk}
 \mathcal{I}^{l}_{n,m,k}
 := \int_p 2\pi\frac{\rmd^l\delta(p^2)}{(\rmd p^2)^l} (p_0)^n |\bp|^{m-n}\frac{\rmd^k f_{\zero\mathrm{mat}}}{\rmd p_0^k} .
\end{equation}
In order to handle the derivative on $\delta(p^2)$, we use the chain rule; e.g. $\frac{\rmd}{\rmd p^2}\delta(p^2) = \frac{1}{2p_0}\frac{\rmd}{\rmd p_0}\delta(p^2)$.
Then, the integration by parts in terms of $p_0$ removes the derivative on $\delta(p^2)$.
It is worthwhile to notice that this step generates no surface term because of $f_{\zero\mathrm{mat}}(p_0 \to \pm\infty) = 0$.
In Appendix~\ref{app:int}, we show the detailed evaluation.
After this step, the integral $\mathcal{I}^{l}_{n,m,k}$ is written as the linear combination of another integral sequence
\begin{equation}
\label{eq:Jab}
\begin{split}
\mathcal{J}_{m,k}
	&:= \int_0^\infty \rmd p\, p^{m} \frac{\rmd^{m}}{\rmd p^{m}}
	\Bigl [n_F(p-\mu) - (-1)^{a+b} n_F(p+\mu)\Bigr]  .
\end{split}
\end{equation}

There is an important remark about the above computation manner.
In Eq.~\eqref{eq:Il_nmk}, we have only the matter part $f_{\zero\mathrm{mat}}$ since the vacuum part is evaluated with the point-splitting regularization.
In some regularization scheme, it is in principle possible to evaluate Eq.~\eqref{eq:Il_nmk} including the vacuum contribution.
In this case, we replace $f_{\zero\mathrm{mat}}$ with $f_\zero = f_{\zero\mathrm{vac}}+f_{\zero\mathrm{mat}}$ in the integrand and evaluate the integral in the almost same manner.
Only one difference is that we carefully take into account the surface term contributions from the $p_0$-integral.
Such contributions always appear for $k=0$ due to the vacuum contribution at ultraviolet regime: $f_{\zero}(p_0 \to +\infty) = 0$ but $f_{\zero}(p_0 \to -\infty) = -1$.
Although Ref.~\cite{Yang:2020mtz} performs a similar integration by parts, the above surface terms are missing.

\subsection{Vacuum part}
Now we compute the vacuum contribution of Eq.~\eqref{eq:integral_general} with the point-splitting regularization.
What we need to evaluate is
\begin{equation}
\label{eq:K_reg}
 \mathcal{K}_{n}^{\mu_1\cdots\mu_m}(y)
 := \int_p 2\pi\frac{\rmd^n\delta(p^2)}{(\rmd p^2)^n} p^{\mu_1}\cdots p^{\mu_m}
 	\bigl[ -\theta(-p_0) \bigr] \rme^{\rmi p\cdot y/\hbar} .
\end{equation}
It is efficient to first evaluate $\mathcal{K}_1$, $\mathcal{K}_2^{\mu\nu}$ and $\mathcal{K}_3^{\mu\nu\rho\sigma}$, which would lead to the logarithmic ultraviolet divergence without the point-splitting.
After the contour deformation to obtain an integral on the Euclidean momentum phase space, we can evaluate these three integrals.
As shown in Appendix~\ref{app:ps}, the result is as follows:
\begin{equation}
\begin{split}
\label{eq:K_123}
 \mathcal{K}_1(y)
= - \frac{\mathcal{J} (y)}{8\pi^2}  ,\quad
\mathcal{K}_2^{\mu\nu} (y)
= \frac{\mathcal{J} (y)}{16\pi^2} g^{\mu\nu} , \quad
\mathcal{K}_3^{\mu\nu\rho\sigma} (y)
= -\frac{\mathcal{J} (y)}{32\pi^2} 
(g^{\mu\nu}g^{\rho\sigma}+g^{\mu\rho}g^{\nu\sigma}+g^{\mu\sigma}g^{\nu\rho})  ,
\end{split}
\end{equation}
with the regularized integral:
\begin{equation}
\label{eq:J_log}
\mathcal{J} (y)
 := \int_0^{y^{-1}} \frac{\rmd p}{p} .
\end{equation}
We again emphasize that the infrared divergence at $p\sim 0$ are completely canceled by those of the matter part.

All other types of integrals in Eq.~\eqref{eq:K_reg} are generated by the derivative of Eq.~\eqref{eq:K_123} with respect to $y^\mu$.
It is important to remind then that in the point-splitting regularization, we take the limit of $y\to 0$ symmetrically at the end of evaluation, as follows~\cite{Peskin:1995ev}:
\begin{equation}
 \underset{y\to 0}{\mathrm{symm\,lim}}\, \frac{y^\mu}{y^2} = 0  ,
 \qquad
 \underset{y\to 0}{\mathrm{symm\,lim}}\, \frac{y^\mu y^\nu}{y^2} = \frac{g^{\mu\nu}}{4}  .
\end{equation}
Thanks to the first equation, for example, we readily find $\mathcal{K}^\mu_1 = -\rmi\hbar \partial^\mu_y\mathcal{K}_1 \propto y^\mu/y^2 \to 0$ in this limit.
Eventually, the integrals~\eqref{eq:K_reg} other than the three in Eq.~\eqref{eq:K_123} vanish in the following section.

\section{Equilibrium transport}\label{sec:JT}

We can now evaluate the charge current~\eqref{eq:J_mu} and the energy-momentum tensor~\eqref{eq:T_munu}, from the momentum integral~\eqref{eq:integral_general}.
It is then convenient to introduce the following four-vector fields:
\begin{equation}
\begin{split}
\label{eq:Fomega}
 &B^\mu := \tF^{\mu\nu} \xi_\nu ,\quad
 E^\mu := F^{\mu\nu}\xi_\nu , \\
 &\omega^\mu := \omega^{\mu\nu}\xi_\nu= \frac{1}{2}\beta^{-1}\varepsilon^{\mu\nu\rho\sigma}\xi_\nu\partial_\rho \beta_\sigma ,\quad
 a^\mu:=\beta^{-1}\xi_\nu\partial^\mu\beta^\nu = \beta^{-1}\partial^\mu \beta  .
\end{split}
\end{equation}
Hereafter, we focus on the equilibrium cases described by either the condition~\eqref{eq:equilibrium1} or~\eqref{eq:equilibrium2}, on top of those in Eq.~\eqref{eq:equilibrium_f0}.
Therefore, in the following analysis, either $a_\mu$ or $\partial_\mu F_{\mu\nu}$ should vanish depending on the choice of Eq.~\eqref{eq:equilibrium1} or~\eqref{eq:equilibrium2}.

The classical and the first-order contributions can be evaluated with the integral formulas in Appendix~\ref{app:int}.
As derived in many literatures, we get~\cite{Chen:2015gta}:
\begin{equation}
\begin{split}
J^\mu_\one &= \frac{\mu}{4\pi^2} B^\mu + \biggl(\frac{\mu^2}{4\pi^2}+\frac{T^2}{12}\biggr) \omega^\mu ,\\
 T^{\mu\nu}_\one &= \biggl(\frac{\mu^2}{4\pi^2} + \frac{T^2}{12}\biggr) B^{(\mu}\xi^{\nu)} + \biggl(\frac{\mu^3}{3\pi^2}+\frac{\mu T^2}{3}\biggr) \omega^{(\mu}\xi^{\nu)}  ,
\end{split}
\end{equation}
which represent the chiral magnetic effect~\cite{Vilenkin:1980fu,Nielsen:1983rb,Fukushima:2008xe} and the chiral vortical effect~\cite{Vilenkin:1979ui,Son:2009tf,Landsteiner:2011cp}.

For the nonlinear-order contributions to Eqs.~\eqref{eq:J_mu} and~\eqref{eq:T_munu}, we differently evaluate the matter and vacuum part, with the help of the integral formulas in Appendices.~\ref{app:int} and~\ref{app:ps}, respectively.
Since $\calR^\mu_\two$ is decomposed into the four different pieces~\eqref{eq:R_delF}-\eqref{eq:R_oo}, so are the corresponding charge current $J^\mu_\two$ and energy-momentum tensor $T^{\mu\nu}_\two$.
The resulting expressions are as follows:
\begin{equation}
\begin{split}
\label{eq:J2}
& J_{(\partial F)}^\mu
=-\frac{\mathcal{J}_{-1,0}-\mathcal{J}}{12\pi^2} \partial_\lambda F^{\mu\lambda}  , \\
& J_{(FF)}^\mu
 = \frac{\mathcal{J}_{-1,1}}{12\pi^2}
 	\biggl[
 		\frac{1}{2} \xi^\mu (E^2+B^2) 
 		+ \varepsilon^{\mu\nu\rho\sigma} \xi_\nu E_\rho B_\sigma
 	\biggr]  ,\\
 & J_{(F\omega)}^\mu
 = \frac{1}{8\pi^2}
 	\Bigl[
 		-\xi^\mu (B\cdot\omega + E\cdot a)
 		+\varepsilon^{\mu\nu\rho\sigma}\xi_\nu B_\rho a_\sigma
 	\Bigr] ,\\
 & J^\mu_{(\omega\omega)}
= - \frac{\mu}{4\pi^2} \xi^\mu (\omega^2 + a^2)  ,
\end{split}
\end{equation}
\begin{equation}
\begin{split}
\label{eq:T2} 
T^{\mu\nu}_{(\partial F)}
&= \frac{\mu}{24\pi^2}
	\biggl[
		-\xi^{(\mu} \partial_\lambda F^{\nu)\lambda}
		+2\xi^\mu\xi^\nu \xi^\lambda \partial^\rho F_{\rho\lambda}
		-g^{\mu\nu}\xi_\lambda \partial_\rho F^{\rho\lambda}
		+\xi_\lambda \partial^{(\mu} F^{\nu)\lambda}
	\biggr]  ,\\
T^{\mu\nu}_{(FF)}
 &= \frac{\mathcal{J}_{-1,0}-\mathcal{J}}{12\pi^2}
 	\biggl[
		{F^{\mu}}_{\sigma} F^{\nu\sigma}
		- \frac{1}{4} g^{\mu\nu} F_{\alpha\beta}^2
 	\biggr]  , \\
T^{\mu\nu}_{(F\omega)}
&=\frac{\mu}{8\pi^2}
	\biggl[
		-\xi^\mu \xi^\nu (\omega\cdot B + a\cdot E)
		+ \omega^{(\mu}B^{\nu)}+a^{(\mu} E^{\nu)}
		+2\xi^{(\mu}\varepsilon^{\nu)\rho\sigma\lambda}a_\rho B_\sigma \xi_\lambda
	\biggr]  ,\\
T^{\mu\nu}_{(\omega\omega)}
&=\biggl(
	\frac{\mu^2}{2\pi^2}
	+ \frac{T^2}{6}
  \biggr)
	 	\biggl[			
			\biggl(
				\frac{1}{4} g^{\mu\nu}
				-\xi^\mu\xi^\nu
			\biggr) (\omega^2+a^2)
			+\xi^{(\mu} \varepsilon^{\nu)\rho\sigma\lambda}a_\rho\omega_\sigma \xi_\lambda 
		\biggr] .
\end{split}
\end{equation}
Here, the longitudinal component of the derivative disappears, i.e., $\xi\cdot\partial F_{\mu\nu} =0$, due to the equilibrium condition~\eqref{eq:equilibrium}.
For the energy-momentum tensor, the term with $Q^\mu$ in Eq.~\eqref{eq:T_munu} yields no contribution, as is readily checked.
We again emphasize that either $\partial_\lambda F_{\mu\nu}$ or $a^\mu$ is admitted to survive due to the conditions~\eqref{eq:equilibrium1} and~\eqref{eq:equilibrium2}, respectively.

We should make a comparison with Ref.~\cite{Yang:2020mtz}.
The authors derived almost the same transport as above, except for $J^\mu_{(\partial F)}$, $T^{\mu\nu}_{(\partial F)}$ and $T^{\mu\nu}_{(FF)}$.
The first two were not computed since the authors focused only on constant background fields.
The stark difference from Ref.~\cite{Yang:2020mtz} is found in $T^{\mu\nu}_{(FF)}$.
The two underlying reasons of this difference are elucidated by recalling the arguments in Sec.~\ref{sec:integral}.
First, the authors did not take into account finite surface terms because of the vacuum contribution in Eq.~\eqref{eq:integral_general}.
Second, they implemented dimensional regularization, without caring about the modification on the Clifford algebra in $d\neq 4$ dimensions.
As a result, while our energy-momentum tensor agrees with that from the Euler--Heisenberg effective theory, that derived in Ref.~\cite{Yang:2020mtz} does not (see Sec.\ref{sec:EH}).

An important observation in Eqs.~\eqref{eq:J2} and~\eqref{eq:T2} is the finite contributions from the magneto-vortical terms $J^\mu_{(F\omega)}$ and $T^{\mu\nu}_{(F\omega)}$.
In particular, the charge density $J^0_{(F\omega)}\sim B\cdot\omega$ agree with that derived in Refs.~\cite{Hattori:2016njk,Ebihara:2016fwa,Yang:2020mtz,Lin:2021sjw}.
There is, however, a crucial contrast with them in terms of the derivations.
On the one hand, the above early studies implicitly assume an equilibrium under magnetic field and vorticity, despite the subtlety of this assumption;
the interplay of magnetic field and rotation classically generates an effective electric field, which in general prohibits the equilibration.
On the other hand, our $J^0_{(F\omega)}\sim B\cdot\omega$ is derived from the equilibrium Wigner function, which is determined by the kinetic equation.
We hence verifies the nondissipativeness of the above magneto-vortical effect, based on quantum field theory.
This is one of the main findings in this paper.

However, the above result does not reproduce the induced current $\sim B\cdot \omega B^\mu/|B|$, which is discovered in Ref.~\cite{Hattori:2016njk}.
This is because our classical Wigner function $\calR^\mu_\zero$ is independent of $B^\mu$.
Contrary, if $\calR^\mu_\zero$ depends on $B^\mu$, there emerges $B^\mu/|B|$ as a possible tensorial basis of $\calR^\mu_\two$, similarly to the fluid velocity $\xi^\mu$.
This is in fact the case of the CKT in the strong magnetic field~\cite{Lin:2021sjw}.
Hence, although the magneto-vortical coupling generates both the charge $\sim B\cdot \omega$ and the current $\sim B\cdot \omega B^\mu/|B|$, they are qualitatively different.
Such a difference would be related to their anomalous nature~\cite{Bu:2019qmd}.

Let us argue the conservation laws for the transport in Eqs.~\eqref{eq:J2} and~\eqref{eq:T2}.
One can compute the divergences $\partial_\mu J^\mu_\two$ with the help of Eq.~\eqref{eq:rec_Jmk} and the formulas in Appendix~\ref{app:Fomega}.
We then observe $\partial_\mu J^\mu_{(F\omega)} = \partial_\mu J^\mu_{(\omega\omega)} = \partial_\mu (J^\mu_{(FF)}+J^\mu_{(\partial F)}) = 0$. Therefore, the nonlinear contribution of the charge current is conserved:
\begin{equation}
 \partial_\mu J^\mu_\two = 0 ,
\end{equation}
where we impose $a_\mu \partial_\nu F_{\rho\sigma} = 0$ because of the equilibrium condition~\eqref{eq:equilibrium}.
This relation holds under both the conditions~\eqref{eq:equilibrium1} and~\eqref{eq:equilibrium2}. 
The divergence $\partial_\mu T^{\mu\nu}_\two$ is computed in a similar manner.
We find $\partial_\mu T^{\mu\nu}_{(F\omega)} + F^{\mu\nu}J_\mu^{(F\omega)} = \partial_\mu T^{\mu\nu}_{(\omega \omega)} + F^{\mu\nu}J_\mu^{(\omega\omega)}= 0$, but
\begin{equation}
\begin{split}
&\partial_\mu T^{\mu\nu}_{(FF)} + F^{\mu\nu}(J_\mu^{(FF)} +J_\mu^{(\partial F)})
= \frac{1}{48\pi^2}
	\Bigl[
			a^\nu F_{\alpha\beta} F^{\alpha\beta} 
 			- 4a^\mu F_{\mu\sigma} F^{\nu\sigma}
	\Bigr] ,\\
 &\partial_\mu T^{\mu\nu}_{(\partial F)} 
 =
 \frac{1}{48\pi^2}
	\Bigl[
  		-\xi^{\nu} E_\mu \partial_\lambda F^{\lambda\mu}
		+2\xi_\mu E^\nu \partial_\lambda F^{\lambda\mu}
		-2 \xi_\lambda E_\mu \partial^{(\mu} F^{\nu)\lambda}
	\Bigr]  ,
\end{split}
\end{equation}
where we again drop the product terms $\sim a_\mu \partial_\lambda F_{\nu\rho}$.
Thus, we arrive at
\begin{equation}
 \partial_\mu T^{\mu\nu}_\two + F^{\mu\nu} J_\mu^\two \neq 0  .
\end{equation}
This violation of the translational invariance is a compensation of the point-splitting regularization.

Lastly, we look at the trace of the energy-momentum tensors in Eq.~\eqref{eq:T2}.
We first notice that $T^{\mu\nu}_{(\partial F)}$, $T^{\mu\nu}_{(F\omega)}$ and $T^{\mu\nu}_{(\omega\omega)}$ are traceless irrelevantly to the regularization scheme.
The same is true for $T^{\mu\nu}_{(FF)}$, as long as we utilize the point-splitting regularization.
Eventually, no trace anomaly is reproduced:
\begin{equation}
\label{eq:T_trace}
{T^\mu}_{\mu\two} = 0  .
\end{equation}
This is another compensation of the point-splitting regularization;
the energy-momentum conservation and the tracelessness do not simultaneously hold.
We emphasize that the QED trace anomaly stems from the fermion loop corrections, regardless of whether electromagnetic fields are background or not~\cite{Giannotti:2008cv,Bastianelli:2018osv,Bastianelli:2022hmu};
it is generally inevitable to introduce some regularization scale.
Hence, Eq.~\eqref{eq:T_trace} is just a consequence of our regularization. 

\section{Consistency with Euler--Heisenberg effective theory}\label{sec:EH}
For consistency check, let us make a comparison with the Euler--Heisenberg effective theory, which is described by the following effective Lagrangian:
\begin{equation}
\begin{split}
\label{eq:L_EH}
\mathcal{L}_{\mathrm{EH}}
&=-\mathcal{F}
	-\frac{e^2}{8 \pi^2}\int_{s_0}^{\infty} \frac{\rmd s}{s}\,\rme^{-s m^2} \frac{\operatorname{Re} \cosh \Bigl[\hbar\,es\sqrt{2(\mathcal{F}+\rmi\mathcal{G})}\Bigr]}{\operatorname{Im} \cosh \Bigl[\hbar\,es\sqrt{2(\mathcal{F}+\rmi\mathcal{G})}\Bigr]}\, \mathcal{G} 
\end{split}
\end{equation}
with $\mathcal{F} := F_{\alpha\beta}^2/4$, $\mathcal{G}:= F^{\alpha\beta}\tF_{\alpha\beta}/4$ and $m$ being the fermion mass.
Here $\hbar$ and $e$ are explicitly written.
In the above Lagrangian, we do not put the conventional counterterms $\sim 1/s^3$ and $\sim e^2\mathcal{F}/s$, which accounts for the vacuum energy renormalization and the charge renormalization.
Instead of this minimal subtraction, we introduced the ultraviolet cutoff parameter $s_0$, which plays the similar role to $y^{-1}$ in the point-splitting regularization.
The charge current and the energy-momentum tensor are obtained from the derivative of the corresponding action with respect to gauge field $A_\mu$ and metric tensor $g_{\mu\nu}$, respectively.
We define them as
\begin{equation}
\begin{split}
\label{eq:def_JT_EH}
 J^\mu_{\mathrm{EH}}
 := -\frac{\delta}{\delta A_\mu} \int \rmd^4 x\, (\hbar^{2} \mathcal{L}_{\mathrm{EH}}) ,\qquad
 T^{\mu\nu}_{\mathrm{EH}}
 := \frac{2}{\sqrt{-g}}\frac{\delta}{\delta g^{\mu\nu}}\int \rmd^4 x \sqrt{-g} \,(\hbar^{2} \mathcal{L}_{\mathrm{EH}}) .
 \end{split}
\end{equation}
For the energy-momentum tensor $T^{\mu\nu}_{\mathrm{EH}}$, we utilized the effective action in a general curved spacetime with $g:=\det(g_{\mu\nu})$.
We note that the factor $\hbar^{2}$ is from our convention for the comparison with the CKT analysis;
on top of $\hbar^{-1}$ by definition of action,  the extra $\hbar^3$ is multiplied because we abbreviate the $\hbar^{-3}$ in the momentum phase space, following the usual convention in the CKT.

The above Lagrangian can be expanded in terms of power of $\hbar$.
This is generally written as follows:
\begin{equation}
\label{eq:L_EH_expanded}
\hbar^2\mathcal{L}_\mathrm{EH}
= -\hbar^2\mathcal{F} 
  +\mathcal{L}_\mathrm{EH\zero} 
  + \hbar^2\mathcal{L}_\mathrm{EH\two} 
  + \hbar^4\mathcal{L}_\mathrm{EH(4)}
  +\cdots  .
\end{equation}
For the latter convenience, here we multiplied $\hbar^2$ by both sides.
In Eq.~\eqref{eq:L_EH_expanded}, what we are now interested in is
\begin{equation}
\begin{split}
\label{eq:L_EH2-mass}
 \mathcal{L}_{\mathrm{EH}(2)} 
  = -\frac{e^2}{48\pi^2}F_{\mu \nu}^2 \int_{s_0}^\infty \frac{\rmd s}{s}\, \rme^{-sm^2}
  = -\frac{e^2}{24\pi^2} F_{\mu \nu}^2
  	\int_{0}^{s_0^{-1/2}} \frac{\rmd p}{p}\, \rme^{-m^2/p^2}  .
\end{split}
\end{equation}
The mass parameter $m$ is the convergence factor of the infrared regime at $s\to\infty$ or $p\to 0$.
In our CKT analysis at equilibrium, we do not care about the infrared divergence, thanks to the cancellation by the matter part of $f_\zero$.
For comparison with the CKT, we consider the limit of $m\to 0$%
~\footnote{%
Even if we take the massless limit after performing the integration in $\mathcal{L}_\mathrm{EH\two}$, the logarithmically divergent behavior in terms of $s_0$ is unchanged.
Thus, the order of taking the limit is irrelevant to the present discussion.
}.
By replacing $s_0^{-1/2}$ with $y^{-1}$, we reduce Eq.~\eqref{eq:L_EH2-mass} to
\begin{equation}
\begin{split}
\label{eq:L_EH2}
 \mathcal{L}_{\mathrm{EH}(2)}\bigl|_{m\to 0}
  = -\frac{e^2\mathcal J}{24\pi^2} F_{\mu \nu}^2  ,
\end{split}
\end{equation}
where $\mathcal J$ is given by Eq.~\eqref{eq:J_log}.
Inserting Eq.~\eqref{eq:L_EH2} into Eq.~\eqref{eq:def_JT_EH} and setting $e=1$ as we do in the CKT, we arrive at the following relations:
\begin{equation}
\label{eq:HE-CKT}
\begin{split}
  J^\mu_{\mathrm{EH}\two}\bigl|_{m\to 0} 
  &= \frac{\mathcal{J}}{6\pi^2} \partial_\lambda F^{\mu\lambda} 
  = 2J^\mu_{(\partial F)\,\mathrm{vac}}  ,\\
  T^{\mu\nu}_{\mathrm{EH}\two}\bigl|_{m\to 0} 
  &=-\frac{\mathcal{J}}{6\pi^2}
 	\biggl[
		{F^{\mu}}_{\sigma} F^{\nu\sigma}
		- \frac{1}{4} g^{\mu\nu} F_{\alpha\beta}^2
 	\biggr]
  =2T^{\mu\nu}_{(FF)\,\mathrm{vac}}  ,
\end{split}
\end{equation}
where `$\mathrm{vac}$' denotes the vacuum contribution.
The factor $2$ on the right-hand sides is understood as the degrees of freedom of chirality.
These relations guarantee the correctness of $J^\mu_{(\partial F)}$ and $T^{\mu\nu}_{(FF)}$ in Eqs.~\eqref{eq:J2} and~\eqref{eq:T2}.
We note that the matter part and the vortical terms in Eqs.~\eqref{eq:J2} and~\eqref{eq:T2} are not included here, as they are not enclosed in Eq.~\eqref{eq:L_EH}.

There is an important remark about the spacetime-dependence of electromagnetic fields.
Since the original Euler--Heisenberg effective theory is for a constant $F_{\mu\nu}$, one might be skeptical that the above comparison is meaningful for $J^\mu_\mathrm{EH\two} \sim \partial_\lambda F^{\mu\lambda}$.
If we take into account the coordinate-dependence of $F_{\mu\nu}$, the effective Lagrangian acquires the derivative corrections.
However, the leading derivative correction is of $O\bigl((\partial F)^2\bigr)$ or of $O\bigl(F\partial^2 F\bigr)$~\cite{Lee:1989vh,Gusynin:1995bc}.
In the power counting of $\hbar$, such a term is the fourth-order term, as it contains four derivatives of gauge field.
For this reason, even when $F_{\mu\nu}$ is spacetime-dependent, the Lagrangian $\mathcal{L}_\mathrm{EH\two}$ is unmodified and thus so is Eq.~\eqref{eq:HE-CKT}.
Therefore, we conclude that the nonlinear CKT is consistent with the Euler--Heisenberg effective theory.

The Euler--Heisenberg effective theory also reveals underlying physics of the logarithmic behavior of $\mathcal J$ in Eqs.~\eqref{eq:J2} and~\eqref{eq:T2}.
To illustrate it, we write the Lagrangian~\eqref{eq:L_EH} in the following form:
\begin{equation}
\label{eq:L_EH_log}
 \hbar^2\mathcal{L}_\mathrm{EH} 
= -\frac{\hbar^2}{4}F_{\mu\nu}^2 
	\biggl(
		1 
		+ \frac{e^2}{12\pi^2} \log\frac{s_0^{-1}}{m^2}
	\biggr)
	+ \mathrm{const.}
	+O(\hbar^4) ,
\end{equation}
where the constant is the term without $F_{\mu\nu}$.
The logarithmic behavior is the same as that found in the vacuum polarization of QED.
From the above Lagrangian, hence, we read off the effective charge $e^2_\mathrm{eff}(M) := e^2 (1 + \frac{e^2}{12\pi^2} \log\frac{M^2}{m^2})$, and the $\beta$-function $\beta (e_\mathrm{eff}) := M\rmd e_\mathrm{eff}(M)/\rmd M = e^3_\mathrm{eff}(M)/(12\pi^2)$~\cite{Schwartz:2014sze}.
Equation~\eqref{eq:HE-CKT} shows that this characteristic of the charge renormalization is inherited not only in the Euler--Heisenberg theory, but also in the nonlinear CKT through the same logarithm $\mathcal J \sim\log y^{-1}$.
At the same time, in spite of Eq.~\eqref{eq:T_trace}, we find that the logarithm $\mathcal J$ in Eq.~\eqref{eq:T2} is an indirect evidence of the trace anomaly, which is determined by the QED $\beta$-function.

\section{Summary}\label{sec:summary}
In this paper, we formulated the nonlinear CKT under arbitrary background electromagnetic fields.
We derived the off-equilibrium Wigner function for arbitrary frame vectors.
Imposing the frame-independence of this Wigner function, we identified an equilibrium Wigner function, which solves the kinetic equation.
As an application, we compute the transport phenomena at the equilibrium.
We then found that the charge induced by the interplay of magnetic field and vorticity~\cite{Hattori:2016njk} are permitted at the equilibrium of the nonlinear CKT.
This analysis based on the Wigner function is, to the best of our knowledge, the first field-theoretical verification of the nondissipativeness of the above charge generation.
Besides, as an important finding, we also showed that the nonlinear CKT and the Euler--Heisenberg effective theory share equivalent transport phenomena.
The ultraviolet logarithmic behavior in the nonlinear CKT is not only a kinetic encoding of the charge renormalization but also an indirect signature of the trace anomaly in the kinetic description.

Also, we posed the potential issue that the prominent schemes, i.e., Pauli--Villars regularization and dimensional regularization, are incompatible with the CKT.
The incompatibility of the latter scheme is one reason of the fact that the energy-momentum tensor in Ref.~\cite{Yang:2020mtz} disagree with that derived from the Euler--Heisenberg effective theory.
For this reason, we employed the point-splitting regularization, which is much more compatible with the Wigner function but cannot directly reproduce the trace anomaly.
For the complete reproduction of the trace anomaly in the kinetic description, we should find out an appropriate regularization in CKT, or rely on frameworks other than the CKT.
For the latter option, the kinetic theory of massive fermions involving the $O(\hbar^2)$ correction is one of the candidates, since Pauli--Villars regularization could be applicable.

Several potential developments are invoked from the nonlinear CKT.
First, the nonlinear transport phenomena is one of the pivotal research fields in condensed matter physics~\cite{boyd2020nonlinear}.
Also, the merit of the nonlinear CKT could be found in, for instance, the so-called nonlinear Hall effect~\cite{PhysRevLett.115.216806,du2021nonlinear}, which originates from the Berry curvature dipole.
In the nonlinear CKT, such contribution would be hidden (see also Refs.~\cite{PhysRevLett.112.166601,PhysRevB.91.214405,Gorbar:2017cwv}, which argue nonlinear corrections of the Berry curvature).
Besides, it is straightforward but complicated to extend the present nonlinear CKT to the collisional case by starting from the Kadanoff--Baym equation with fermionic self-energy~\cite{kadanoff1962quantum,Blaizot:2001nr}.
In the nonlinear CKT, it is also interesting to take into account dynamical gauge fields, which bring the chiral plasma instabilities~\cite{Akamatsu:2013pjd}.
These applications will be discussed elsewhere.

\begin{acknowledgements}
The author thanks Yoshimasa~Hidaka for giving valuable comments.
\end{acknowledgements}

\appendix

\section{Charge current, energy-momentum tensor and spin tensor at $O(\hbar^2)$}\label{app:JTS}
In this Appendix, we derive the charge current and energy-momentum tensor with the Wigner function.
The former for the right-handed massless fermions is defined as
\begin{equation}
\begin{split}
 J^\mu(x,y)
 := \mathrm{tr}\Bigl\langle \bar\psi_+ \gamma^\mu P_\mathrm{R} \psi_- \Bigr\rangle  ,
\end{split}
\end{equation}
where we define $P_\mathrm{R} := \frac{1}{2}(1+\gamma^5)$, $O_- := \rme^{-y\cdot D/2} O(x)$ and $O_+ := O(x) \rme^{y\cdot \overleftarrow{D}/2}$ with $D_\mu := \partial_\mu + \rmi A_\mu/\hbar$, $\overleftarrow{D}_\mu := \overleftarrow{\partial}_\mu - \rmi A_\mu/\hbar$.
Here, the operators $ \rme^{-y\cdot D/2}$ and $\rme^{y\cdot \overleftarrow{D}/2}$ represent the covariant translation, and thus their insertion is equivalent to enclosing the Wilson line~\cite{Elze:1986qd}.
In the $y\to 0$ limit, the above current is reduced to the usual definition in quantum field theory.
Let us here recall that the Wigner function is defined as
\begin{equation}
 \calR^\mu(x,p) := \frac{1}{2}\mathrm{tr}
 	\Bigl[
 		\gamma^\mu P_\mathrm{R} W(x,p)
 	\Bigr]  ,
 	\quad
 W_{ab}(x,p) := \int_y \rme^{-\rmi p\cdot y/\hbar}\, \mathrm{tr}\Bigl\langle (\bar\psi_+)_b (\psi_{-})_a \Bigr\rangle  .
\end{equation}
Then, performing the inverse Wigner transformation, we write the above current as Eq.~\eqref{eq:J_mu}:
\begin{equation}
\label{eq:Jxy_app}
 J^\mu(x,y)
 = 2\int_p \rme^{\rmi p\cdot y/\hbar} \calR^\mu(x,p)  .
\end{equation}
For the spin tensor, the inverse Wigner transformation of the standard field-theoretical definition yields Eq.~\eqref{eq:S_munurho}:
\begin{equation}
\begin{split}
 S^{\mu\nu\rho}(x,y)
&:=\frac{\hbar}{4}\,\mathrm{tr}\Bigl\langle
	\bar\psi_+ \bigl\{\gamma^\mu,\sigma^{\nu\rho}\bigr\} P_\mathrm{R} \psi_-
	\Bigr\rangle 
=-2\hbar\varepsilon^{\mu\nu\rho\sigma}\int_p \rme^{\rmi p\cdot y/\hbar} \calR_\sigma(x,p) 
\end{split}
\end{equation}
with $\sigma^{\mu\nu} = \frac{\rmi}{2}[\gamma^\mu,\gamma^\nu]$.

Let us derive the kinetic expression of the energy-momentum tensor.
Unlike the charge current and spin tensor, the definition of the energy-momentum tensor is ambiguous due to the derivative operator.
We here employ the canonical energy-momentum tensor defined as follows:
\begin{equation}
\begin{split}
\label{eq:Tcan}
 T_\text{can}^{\mu\nu} (x,y)
 &:=\frac{\rmi \hbar}{2} (t^{\mu\nu}-g^{\mu\nu}{t^\lambda}_\lambda) ,\quad
 t^{\mu\nu}
 =
 \mathrm{tr}
 \Bigl\langle
 	\bar\psi_+ \gamma^{\mu} P_\mathrm{R} (D^{\nu}\psi)_- 
 	-(\bar\psi\overleftarrow{D}^{\nu})_+\gamma^{\mu} P_\mathrm{R}\psi_- 
 \Bigr\rangle .
\end{split}
\end{equation}
Note that $(D^{\nu}\psi)_- = \rme^{-y\cdot D/2} D_\mu\psi$ is inequivalent to $D_\mu \psi_- = D_\mu \rme^{-y\cdot D/2} \psi$ when electromagnetic fields are spacetime-dependence [see Eq.~\eqref{eq:FG-identity}].
In the limit of $y\to 0$, this definition is consistent with the classical canonical momentum tensor
\begin{equation}
\Theta^{\mu\nu}(x)
= \partial^\mu \psi\frac{\partial\mathcal{L}}{\partial\partial_\nu \psi}
  +\partial^\mu\bar{\psi}\frac{\partial\mathcal{L}}{\partial\partial_\nu \bar\psi}
  - g^{\mu\nu}\mathcal{L}
\end{equation}
with $\mathcal{L} = \frac{\rmi\hbar}{2}\Bigl[\bar\psi(x) \gamma^{\lambda} P_\mathrm{R} D_\lambda\psi(x)-\bar\psi(x)\overleftarrow{D}_\lambda \gamma^\lambda P_\mathrm{R}\psi(x)\Bigr]$.
In Eq.~\eqref{eq:Tcan}, the last term with $g^{\mu\nu}$ vanishes due to the Dirac equation.
To reduce the first two terms, we prepare the following identities:
\begin{equation}
\begin{split}
\label{eq:FG-identity}
 D_\mu \rme^{y\cdot D}\psi(x)
 &= \biggl[
 		\rme^{y\cdot D} D_\mu + \frac{\rmi y^\lambda}{\hbar} \mathcal{F}_{\mu\lambda}(x,y)\rme^{y\cdot D}
  \biggr]
   \psi(x)  ,\\
 \partial_\mu^y  \rme^{y\cdot D} \psi (x) 
 &= \biggl[ D_\mu \rme^{y\cdot D}  - \frac{\rmi y^\lambda}{\hbar} \mathcal{G}_{\mu\lambda}(x,y) \rme^{y\cdot D} \biggr] \psi (x) 
\end{split}
\end{equation}
with
\begin{equation}
 \mathcal{F}_{\mu\lambda}(x,y)
 =\sum_{n=0}^\infty \frac{(y\cdot\partial)^n}{(n+1)!} F_{\mu\lambda}(x)  ,\quad \mathcal{G}_{\mu\lambda}(x,y)
 =\sum_{n=0}^\infty \frac{(y\cdot\partial)^n}{(n+2)!} F_{\mu\lambda} (x)  .
\end{equation}
These are derived from $\rme^{Y} X \rme^{-Y} = \rme^{\mathcal{C}(Y)} X$ with $\mathcal{C}(Y) X:=[Y,X]$.
Performing the inverse Wigner transformation, we rewrite Eq.~\eqref{eq:Tcan} as
\begin{equation}
\begin{split}
 T^{\mu\nu}_\text{can}(x,y) 
 &= 
 	\biggl[
 		-\rmi \hbar\partial^\nu_y 
 		+ \frac{1}{12}y\cdot\partial F^{\nu\lambda}y_\lambda
 	\biggr]
 		\mathrm{tr}\Bigl\langle\bar\psi_+ \gamma^{\mu} P_\mathrm{R} \psi_- \Bigr\rangle \\
 &= 2\int_p \rme^{ip\cdot y/\hbar} 
 	p^\nu \calR^\mu (x,p)
 	+ \frac{1}{12}y\cdot\partial F^{\nu\lambda}y_\lambda
 	\cdot 2\int_p \rme^{ip\cdot y/\hbar}
 		\calR^\mu (x,p)  \\
\end{split}
\end{equation}
up to $O(\hbar^2)$.
In the second line, we need carefully to perform the integral by parts because the surface terms in general are generated.
At least at the equilibrium described by Eq.~\eqref{eq:equilibrium_f0}, however, we can show that no surface term appears, as follows.
The second term can be decomposed into the contributions from the vacuum and the matter parts, namely, $f_{\zero\mathrm{vac}}(p_0)= -\theta(-p_0)$ and $f_{\zero\mathrm{mat}}(p_0)=\theta(p_0)n_F(p_0-\mu) + \theta(-p_0)n_F(-p_0+\mu)$.
The former should be proportional to $ y^\lambda y^\rho y^{-3}$ for the dimensional reason, and thus vanishes in the symmetric limit of $y\to 0$.
The latter yields no surface term because of $f_{\zero\mathrm{mat}}(p_0)\delta(p^2) \to 0$ for $p_\mu \to 0$.
Therefore, at the equilibrium, performing the integral by parts leads to
\begin{equation}
\begin{split}
 T^{\mu\nu}_\text{can}(x,y) 
 & = 2\int_p \rme^{\rmi p\cdot y/\hbar} (p^\nu + \hbar^2 Q^\nu )\calR^\mu (x,p)  .
\end{split}
\end{equation}
Its symmetric part is given by Eq.~\eqref{eq:T_munu}.

\section{Solution at $O(\hbar^2)$}\label{app:2nd_solution}
In this Appendix, we derive the second-order solution $\calR^\mu_\two$.
The basic step of the following calculation is parallel to Ref.~\cite{Hayata:2020sqz}.
Inserting $\calR_\zero^\mu$ and $\calR_\one^\mu$ into Eq.~\eqref{eq:R4}, we find
\begin{equation}
\label{eq:Rmu2_app}
 \calR_\mu^\two
 = 2\pi\delta(p^2) \widetilde{\calR}_\mu^\two 
    + \frac{2\pi}{p^2} 
		\biggl[
      		- p_\mu Q\cdot p f_\zero
      		+ p^\nu\mathcal{D}_{\mu\nu}
      		- p^\nu \tF_{\mu\nu} f_\one
		\biggr] \delta(p^2)  ,
\end{equation}
with $\widetilde{\calR}_\mu^\two$ satisfying $p^2 \delta(p^2)\widetilde{\calR}^\mu_\two  = \delta(p^2) p\cdot\widetilde{\calR}_\two = 0$.
Here, we have introduced
\begin{equation}
 \mathcal{D}_{\mu\nu} \delta(p^2)
 :=  \frac{1}{2}\varepsilon_{\mu\nu\rho\sigma} \Delta^\rho
  		\biggl(\Sigma^{\sigma\lambda}_n (\Delta_\lambda f_\zero) - \frac{1}{p^2}\tF^{\sigma\lambda}  p_\lambda f_\zero \biggr)\delta(p^2)
  	+ 2 Q_{[\mu} p_{\nu]} f_\zero \delta(p^2)  ,
\end{equation}
where $(\Delta_\lambda f_\zero)$ represents the derivative operation acting only on $f_\zero$, but others operate on all on the right.
For this $\calR^\mu_\two$, Eq.~\eqref{eq:R3} yields
\begin{equation}
 \begin{split}
 \label{eq:tcalR_2}
   \widetilde{\calR}_\mu^\two \delta(p^2)
  	& = \delta(p^2) 
  		  \Bigl[
 			 p_\mu f_\two + \Sigma_{\mu\nu}^u \Delta^\nu f_\one
 		  \Bigr] 
  		 + \frac{1}{p^2} \varepsilon^{\alpha\beta\gamma\nu} \Sigma_{\mu\nu}^u
  	 			p_\alpha \mathcal{D}_{\beta\gamma}\delta(p^2)   ,
 \end{split}
\end{equation}
where we introduce a vector $u^\nu$, and define $f_\two := u\cdot \widetilde{\calR}_\two /(p\cdot u) $ and $\Sigma_u^{\mu\nu} := \varepsilon^{\mu\nu\rho\sigma} p_\rho u_\sigma/(2p\cdot u)$, similarly to Eq.~\eqref{eq:f1_Sigman}.
The last term with complicated structure due to the Levi-Civita symbols can be reduced with the Schouten identity: $\varepsilon^{\mu\nu\rho\sigma}p^{\lambda} + \varepsilon^{\nu\rho\sigma\lambda}p^{\mu} + \varepsilon^{\rho\sigma\lambda\mu}p^\nu+\varepsilon^{\sigma\lambda\mu\nu}p^\rho+\varepsilon^{\lambda\mu\nu\rho}p^\sigma = 0$ and
\begin{equation}
\label{eq:Sigma_p}
\begin{split}
 &\Sigma_n^{\lambda[\mu} p^{\nu]}
 = 
 -\frac{1}{2} \Sigma_n^{\mu\nu} p^\lambda 
 -\frac{1}{4} \varepsilon^{\mu\nu\lambda\rho}p_\rho
 +\frac{1}{4} \varepsilon^{\mu\nu\lambda\rho} \frac{n_\rho p^2}{p\cdot n}  .
\end{split}
\end{equation}
In the CKT at $O(\hbar^2)$, Eq.~\eqref{eq:Sigma_p} is quite helpful in the sense that the frame-independent part and the $p^2$ term can be extracted.
After straightforward computation with these relations, we arrive at
\begin{equation}
\label{eq:calD_tenor}
\begin{split}
 \frac{1}{p^2}\varepsilon^{\alpha\beta\gamma\nu}\Sigma_{\mu\nu}^u
 	p_\alpha \mathcal{D}_{\beta\gamma} \delta(p^2)  
 &= - \delta(p^2)\Sigma_{\mu\nu}^u\varepsilon^{\nu\rho\sigma\lambda}
		\Delta_{\rho} \frac{n_\sigma}{2 p\cdot n}\Delta_\lambda f_\zero \\
 &\quad
 	+ \frac{\delta(p^2)}{p^2}\Sigma_{\mu\nu}^u 
		\biggl[
			\Delta_\alpha\Sigma^{\alpha\nu}_n
			+\frac{n_\alpha}{p\cdot n}\tF^{\alpha\nu}
			+\frac{1}{p^2}\tF^{\nu\lambda}p_\lambda
		\biggr]p\cdot\Delta f_\zero  .
\end{split}
\end{equation}
It should be mentioned that the singular factors $(p^2)^{-1}$ and $(p^2)^{-2}$ in Eq.~\eqref{eq:calD_tenor} does not conflict with the nonsingular condition $\delta(p^2) p^2\widetilde{\calR}^\mu_\two = 0$.
One can show this by noting $(p^2)^{-n}\delta(p^2)p\cdot\Delta f_\zero \neq 0$  for $n\geq 1$ but $\delta(p^2)p\cdot\Delta f_\zero = 0$, which follows from the classical kinetic equation~\eqref{eq:R1}.
Plugging Eqs.~\eqref{eq:tcalR_2} and~\eqref{eq:calD_tenor} into Eq.~\eqref{eq:Rmu2_app}, and proceeding computation, we obtain the second-order solution $\calR^\mu_\two$ in Eq.~\eqref{eq:Rmu2}.

\section{Integral formulas for matter contribution}\label{app:int}
In this Appendix, we derive the integral formulas for the matter contribution.
At equilibrium, the matter contribution in Eq.~\eqref{eq:integral_general} is the following form:
\begin{equation}
 \int_p 2\pi\frac{\delta(p^2)}{(p^2)^l} p^{\mu_1}\cdots p^{\mu_j} \frac{\rmd^k f_{\zero \mathrm{mat}}}{\rmd p_0^k} 
\end{equation}
with $f_{\zero \mathrm{mat}} = \theta(p_0) n_F(p_0-\mu) + \theta(-p_0) n_F(-p_0+\mu)$ and $n_F(x) = (\rme^{\beta x} + 1)^{-1}$.
In the integrands, we can implement the following replacement:
\begin{equation}
\begin{split}
\label{eq:p_replacement}
 p_{\alpha}
	& \to p_{0}\xi_{\alpha}  , \\
 p_{\alpha}p_{\beta} 
	& \to (p_{0})^{2}\xi_{\alpha}\xi_{\beta} 
		+ \frac{\bp^{2}}{3}\Delta_{\alpha\beta},\\
 p_\alpha p_\beta p_\gamma
 &\to (p_0)^3\xi_\alpha\xi_\beta \xi_\gamma
 	 + \frac{p_0\bp^2}{3}
 	 	(\xi_\alpha\Delta_{\beta\gamma}
 	 	 + \xi_\beta\Delta_{\gamma\alpha}
 	 	 + \xi_\gamma\Delta_{\alpha\beta}), \\
p_{\alpha}p_{\beta}p_{\gamma}p_{\delta} 
	& \to (p_{0})^{4}\xi_{\alpha}\xi_{\beta}\xi_{\gamma}\xi_{\delta} \\
	& \quad
		 + \frac{(p_{0})^{2}\bp^{2}}{3}
		 	(\xi_{\alpha}\xi_{\beta}\Delta_{\gamma\delta} 
		 	 + \xi_{\alpha}\xi_{\gamma}\Delta_{\beta\delta} 
		 	 + \xi_{\alpha}\xi_{\delta}\Delta_{\beta\gamma}
		 	 + \xi_{\beta}\xi_{\gamma}\Delta_{\alpha\delta} 
		 	 + \xi_{\beta}\xi_{\delta}\Delta_{\alpha\gamma}
		 	 + \xi_{\gamma}\xi_{\delta}\Delta_{\alpha\beta}) \\
	& \quad
		 + \frac{|\bp|^4}{15} 
		 	(\Delta_{\alpha\beta}\Delta_{\gamma\delta}
		 	 + \Delta_{\alpha\gamma}\Delta_{\beta\delta}
		 	 + \Delta_{\alpha\delta}\Delta_{\beta\gamma}),
\end{split}
\end{equation}
with $\xi^\mu := (1,\boldsymbol{0})$ and the transverse projector $\Delta^{\mu\nu}:=\xi^{\mu}\xi^{\nu}-g^{\mu\nu}$.
Then, the above integral is represented as a linear combination of
\begin{equation}
\label{eq:Il_nmk_app}
 \mathcal{I}^{l}_{n,m,k}
 = \int_p 2\pi\frac{\rmd^l\delta(p^2)}{(\rmd p^2)^l} F_{n,m,k} ,
 \quad
  F_{n,m,k}:=(p_0)^n |\bp|^{n-m}\frac{\rmd^k f_{\zero\mathrm{mat}}}{\rmd p_0^k} ,
\end{equation}
with $\int_p=\int \frac{\rmd^4 p}{(2\pi)^4}$.
We start from
\begin{equation}
\label{eq:delta_pm}
 \delta(p^2) 
 = \frac{1}{2|\bp|} (\delta_+ + \delta_- )  ,
 \quad
 \delta_{\pm}:=\delta(p_0\mp|\bp|)  .
\end{equation}
Then the first, second, and third derivatives are computed as
\begin{equation}
\begin{split}
\label{eq:delta_pm_prime}
 \delta'(p^2)
 &=\frac{1}{4p_0|\bp|}(\delta'_+ + \delta'_- )  , \\
 \delta''(p^2)
 &=-\frac{1}{8p_0^3|\bp|}(\delta'_+ + \delta'_- )
 	+ \frac{1}{8p_0^2|\bp|}(\delta''_+ + \delta''_- )  , \\
 \delta'''(p^2)
 &=\frac{3}{16p_0^5|\bp|}(\delta'_+ + \delta'_- )
 	-\frac{3}{16p_0^4|\bp|}(\delta''_+ + \delta''_- )
 	+ \frac{1}{16p_0^3|\bp|}(\delta'''_+ + \delta'''_- )  .
\end{split}
\end{equation}
where the primes on $\delta_\pm$ denote the derivative with respect to $p_0$.
For $l=0$, we readily compute Eq.~\eqref{eq:Il_nmk_app} by using Eq.~\eqref{eq:delta_pm}. 

For $l\geq 1$, performing the integration by parts, we can replace $\rmd/\rmd p_0$ in Eq.~\eqref{eq:delta_pm_prime} with that on $F_{n,m,k}$.
For instance, the integral for $l=1$ reads
\begin{equation}
\begin{split}
\label{eq:I1_nmk}
 \mathcal{I}^1_{n,m,k}
 &= - \int_\bp \frac{1}{4|\bp|} \int \rmd p_0  (\delta_+ + \delta_-) \frac{\rmd}{\rmd p_0}\frac{F_{n,m,k}}{p_0}
\end{split}
\end{equation}
with $\int_\bp = \int \frac{\rmd^3 p}{(2\pi)^3}$.
In a similar manner, we obtain
\begin{equation}
\begin{split}
\label{eq:I2_nmk}
 \mathcal{I}^2_{n,m,k}
 &= \int_\bp \frac{1}{8|\bp|} \int \rmd p_0 (\delta_+ + \delta_-)
 	\biggl[
 		 \frac{\rmd}{\rmd p_0}\frac{F_{n,m,k}}{p_0^3} 
 		+ \frac{\rmd^2}{\rmd p_0^2} \frac{F_{n,m,k}}{p_0^2}
 	\biggr]  ,
\end{split}
\end{equation}
\begin{equation}
\begin{split}
\label{eq:I3_nmk}
 \mathcal{I}^3_{n,m,k}
 &= -\int_\bp \frac{1}{16|\bp|} \int \rmd p_0 (\delta_+ + \delta_-) 
 	\biggl[
 		\frac{\rmd}{\rmd p_0}\frac{3F_{n,m,k}}{p_0^5} 
 		+ \frac{\rmd^2}{\rmd p_0^2}\frac{3 F_{n,m,k}}{p_0^4} 
 		+ \frac{\rmd^3}{\rmd p_0^3} \frac{F_{n,m,k}}{p_0^3}
 	\biggr]  .
\end{split}
\end{equation}
Carrying out the momentum integration in Eqs.~\eqref{eq:I1_nmk}, \eqref{eq:I2_nmk} and~\eqref{eq:I3_nmk}, we finally derive
\begin{eqnarray}
\mathcal{I}_{n,m,k}^0 &=& \frac{1}{4\pi^2} \mathcal{J}_{m+1,k}  ,\\
\label{eq:I1_nmk_app}
\mathcal{I}_{n,m,k}^1 &=& \frac{-1}{8\pi^2} \biggl[(n-1)\mathcal{J}_{m-1,k} + \mathcal{J}_{m,k+1} \biggr]  ,\\
\label{eq:I2_nmk_app}
\mathcal{I}_{n,m,k}^2 &=& \frac{1}{16\pi^2} \biggl[(n-1)(n-3) \mathcal{J}_{m-3,k} + (2n-3)\mathcal{J}_{m-2,k+1} 
+\mathcal{J}_{m-1,k+2} \biggr] , \\
\label{eq:I3_nmk_app}
\mathcal{I}_{n,m,k}^3 &=& \frac{-1}{32\pi^2} \biggl[(n-1)(n-3)(n-5)\mathcal{J}_{m-5,k}
+ 3(n^2-5n+5)\mathcal{J}_{m-4,k+1}\nonumber \\
&&\qquad\quad 
 +3(n-2)\mathcal{J}_{m-3,k+2} +\mathcal{J}_{m-2,k+3} \biggr]  ,
\end{eqnarray}
where the integral sequence $\mathcal{J}_{m,k}$ is given by
\begin{equation}
\mathcal{J}_{m,k} := \int_0^\infty \rmd p\, p^{m} \frac{\rmd^{k}}{\rmd p^{k}}
	\Bigl [n_F(p-\mu) - (-1)^{m+k} n_F(p+\mu) \Bigr]  .
\end{equation}
One can show the following recursion equations for $m\geq 0$ and $k\geq 0$:
\begin{equation}
\label{eq:rec_Jmk}
 \mathcal{J}_{m+1,k+1} = -(m+1)\mathcal{J}_{m,k} ,\quad
 \partial_\mu \mathcal{J}_{m,k}
 =E_\mu \mathcal{J}_{m,k+1}
 +a_\mu (\mathcal{J}_{m+1,k+1}+k\mathcal{J}_{m,k}) .
\end{equation}
The former is useful to reduce Eqs.~\eqref{eq:I1_nmk_app}-\eqref{eq:I3_nmk_app}, and the latter is helpful to compute the divergence of $J^\mu$ and $T^{\mu\nu}$.

\section{Point-splitting regularization}\label{app:ps}

In this Appendix, we demonstrate the evaluation of $\mathcal{K}_1$, $\mathcal{K}^{\mu\nu}_2$ and $\mathcal{K}^{\mu\nu\rho\sigma}_3$, where
\begin{equation}
\label{eq:K_reg_app}
 \mathcal{K}_{n}^{\mu_1\cdots\mu_m}(y)
 := \int_p 2\pi\frac{\rmd^n\delta(p^2)}{(\rmd p^2)^n} p^{\mu_1}\cdots p^{\mu_m}
 	\bigl[ -\theta(-p_0) \bigr] \rme^{\rmi p\cdot y/\hbar} .
\end{equation}
As usual, the point-splitting regularization is implemented with the Euclidean momentum integral.
For the above integral, the simple Wick rotation with $p_0\to -\rmi p_4$ cannot be admitted due to the delta function and step function, which are defined on real space.
For this reason, we first write them as
\begin{equation}
\delta(x) = \frac{1}{\pi}{\rm Im}\frac{1}{x-\rmi \epsilon} ,\quad
\theta(x) = \frac{1}{2\pi\rmi} \int_{-\infty}^\infty \rmd\tau \frac{\rme^{\rmi x\tau}}{\tau-\rmi \eta} 
\end{equation}
with positive infinitesimals $\epsilon$ and $\eta$.

Let us first compute $\mathcal{K}_1(y)$, which can be expressed as
\begin{equation}
\label{eq:K1_app}
 \mathcal{K}_1 (y)
 = \frac{1}{2\rmi}(\mathcal{K}_+ - \mathcal{K}_-)  ,
 \quad
 \mathcal{K}_\pm :=
 \frac{-2}{2\pi \rmi} \int_{-\infty}^{\infty}\frac{\rmd\tau}{\tau + \rmi\eta}\int_p  \frac{\rme^{\rmi p_0(\tau+y_0/\hbar)-\rmi\bp\cdot\boldsymbol{y}/\hbar} }{(p^2\mp \rmi\epsilon)^2}  .
\end{equation}
We can now deform the contours of $p_0$-integral, together with that of $\tau$-integral, along the imaginary axis.
Introducing another positive infinitesimal $\delta$, we compute
\begin{equation}
\begin{split}
\mathcal{K}_{\pm}
&= \frac{-2}{2\pi \rmi} \int_{\rmi\infty+\delta}^{-\rmi\infty+\delta}\frac{\rmd\tau}{\tau + \rmi\eta}
	\int_\bp\int_{\rmi\infty}^{-\rmi\infty}  (\pm 1)\frac{\rmd p_0}{2\pi}  \frac{\rme^{\rmi p_0(\tau+y_0/\hbar)-\rmi\bp\cdot\boldsymbol{y}/\hbar}}{(p^2\mp \rmi\epsilon)^2} \\
&=(\pm 1)\cdot\frac{-2}{2\pi\rmi}\cdot\frac{1}{\rmi} \int_{-\infty}^{\infty}\frac{\rmd\tau_E}{-\rmi\tau_E +\delta + \rmi\eta}\int_\bp \int_{-\infty}^{\infty} \frac{\rmd p_4}{2\pi\rmi}  \frac{\rme^{p_4(-\rmi\tau_E +\delta+y_0/\hbar)-\rmi \bp\cdot\boldsymbol{y}/\hbar}}{(-p^2_E)^2} \\
&=(\pm 1) \cdot\frac{-2}{2\pi\rmi} \cdot\frac{1}{\rmi}\int_{-\infty}^{\infty}\frac{\rmd\tau_E}{\tau_E+\rmi\delta}\int_{p_E}  \frac{\rme^{-\rmi p_4(\tau_E+\rmi\delta)-\rmi p_E\cdot y_E/\hbar}}{(p^2_E)^2} \\
&=(\pm 1)\cdot (-2\rmi) \int_{p_E} \theta(p_4) \frac{\rme^{-\rmi p_E\cdot y_E/\hbar}}{(p^2_E)^2}  ,
\end{split}
\end{equation}
where we denote the Euclidean splitting parameter as $y_E = \sqrt{y_4^2+\boldsymbol{y}^2}$ with $y_4:= iy_0$ and inner product as $p_E\cdot y_E = \boldsymbol{p}\cdot\boldsymbol{y} + p_4 y_4$.
In the following, we suppress the subscript $E$.
Due to this contour deformation, the integral \eqref{eq:K1_app} is represented as the momentum integral in the Euclidean four-dimensional half hypersphere:
\begin{equation}
\begin{split}
\label{eq:calK1_1}
 \mathcal{K}_1(y)
 &= -2\int_{p}\theta(p_4) \frac{\rme^{-\rmi p\cdot y/\hbar}}{p^4}  
 =-\frac{1}{4\pi^2}\int_0^\infty \frac{\rmd p}{p} 
 \int_{p_4>0} \frac{\rmd\Omega}{2\pi^2} \rme^{-\rmi py\cos\omega/\hbar} ,
\end{split}
\end{equation}
where $\omega$ is the angular valuable defined by $p\cdot y = py\cos\omega$.

To proceed, it is useful to introduce two integral sequences.
The first one is
\begin{equation}
 \mathcal{Z}_n (x)
 := \int_0^1 \frac{\rmd \zeta}{\pi/2} \zeta^n \sqrt{1-\zeta^2} \,\rme^{-\rmi x \zeta}  .
\end{equation}
This $\mathcal{Z}_n (x)$ can be written with the Bessel function of the first kind $J_n(x)$ and the Struve function ${\bf H}_n(x)$, as follows:
\begin{equation}
\begin{split}
\mathcal{Z}_1(x)
 &= \frac{2}{3\pi} -\rmi\frac{J_2(x)}{x} - \frac{{\bf H}_2(x)}{x}  ,\\
\mathcal{Z}_2(x)
 &= \frac{-2\rmi x}{15\pi} +\frac{J_2(x)-xJ_3(x)}{x^2}
 +\rmi\frac{-{\bf H}_2(x)+x{\bf H}_3(x)}{x^2}  ,\\
\mathcal{Z}_3(x)
 &= \frac{4}{15\pi} +\rmi\frac{-3J_3(x)+xJ_4(x)}{x^2}
 +\frac{3{\bf H}_3(x)-x{\bf H}_2(x)}{x^2}  ,\\
\mathcal{Z}_4(x)
 &= -\frac{2\rmi x}{21\pi} +\frac{2xJ_4(x)-(x^2-3)J_3(x)}{x^3}
 +\rmi\frac{-2x{\bf H}_4(x)+(x^2-3){\bf H}_3(x)}{x^3}  ,\\
\mathcal{Z}_5(x)
 &= -\frac{2(x^2-8)}{105\pi} 
 +\rmi\frac{(x^2-15)J_4(x)}{x^3}
 +\frac{-(x^2-15){\bf H}_4(x)}{x^3}  .
\end{split}
\end{equation}
One important property of $\mathcal{Z}_n(x)$ is that the integral from $0$ to $\infty$ are analytically evaluated as:
\begin{equation}
\begin{split}
 \int_0^\infty \rmd z\, \mathcal{Z}_1(z) &= -\frac{\rmi}{2}  ,\quad
 \int_0^\infty \rmd z\, \mathcal{Z}_2(z) = -\frac{2\rmi}{3\pi}  ,\quad
 \int_0^\infty \rmd z\, \mathcal{Z}_3(z) = -\frac{\rmi}{8}  ,\\
 \int_0^\infty \rmd z\, \mathcal{Z}_4(z) &= -\frac{4i}{15\pi}  ,\quad
  \int_0^\infty \rmd z\, \mathcal{Z}_5(z) = -\frac{\rmi}{16}  .
\end{split}
\end{equation}
Another property is the recurrence relation
\begin{equation}
 \mathcal{Z}'_n(x) = -\rmi \mathcal{Z}_{n+1}(x) ,
 \quad
 \mathcal{Z}_n(x) = -\rmi \int_\infty^x \rmd z\, \mathcal{Z}_{n+1}(z)  ,
\end{equation}
where the latter follow from $\mathcal{Z}_n(x) \underset{x\to\infty}{\longrightarrow} 0$.

The second useful integral is
\begin{equation}
 \mathcal{A}^{\mu_1\cdots \mu_n} (x) 
 := \int_{p_4>0} \frac{\rmd\Omega}{2\pi^2} \hat{p}^{\mu_1}\cdots\hat{p}^{\mu_n} \rme^{-\rmi x\cos\omega}  .
\end{equation}
This tensor is decomposed into the longitudinal component to $\hat{y}^\mu := y^\mu/y$ and transverse one with the projector $\tilde{\Delta}^{\mu\nu} := \delta^{\mu\nu}-\hat{y}^\mu\hat{y}^\nu$.
The coefficients of them is determined by $\mathcal{Z}_n$, as follows:
\begin{equation}
\begin{split}
 \mathcal{A}   
 &= \mathcal{Z}_0  ,\quad
 \mathcal{A}^\mu
 =\hat{y}^\mu \mathcal{Z}_1 ,\quad
 \mathcal{A}^{\mu\nu}  
 = \hat{y}^\mu\hat{y}^\nu \mathcal{Z}_2
   + \tilde{\Delta}^{\mu\nu} \frac{1}{3}\Bigl[\mathcal{Z}_0-\mathcal{Z}_2\Bigr]  ,\\
 \mathcal{A}^{\mu\nu\rho}  
 & = \hat{y}^\mu\hat{y}^\nu\hat{y}^\rho \mathcal{Z}_3
   + \left(\hat{y}^\mu\tilde{\Delta}^{\nu\rho}+\hat{y}^\nu \tilde{\Delta}^{\rho\mu}+\hat{y}^\rho \tilde{\Delta}^{\mu\nu}\right) \frac{1}{3}\Bigl[ \mathcal{Z}_1 -\mathcal{Z}_3 \Bigr]  ,\\
 \mathcal{A}^{\mu\nu\rho\sigma} 
 & =\hat{y}^\mu \hat{y}^\nu \hat{y}^\rho \hat{y}^\sigma
 \mathcal{Z}_4  \\
 &\quad 
 +\left(\tilde{\Delta}^{\mu\nu} \hat{y}^\rho \hat{y}^\sigma+\tilde{\Delta}^{\nu\rho} \hat{y}^\sigma \hat{y}^\mu+\tilde{\Delta}^{\rho\sigma} \hat{y}^\mu \hat{y}^\nu+\tilde{\Delta}^{\mu\sigma} \hat{y}^\nu \hat{y}^\rho+\tilde{\Delta}^{\mu\rho} \hat{y}^\nu \hat{y}^\sigma+\tilde{\Delta}^{\nu\sigma} \hat{y}^\mu \hat{y}^\rho\right)
 \frac{1}{3}\Bigl[ \mathcal{Z}_2-\mathcal{Z}_4 \Bigr]  ,\\
&\quad
 + \left(\tilde{\Delta}^{\mu\nu} \tilde{\Delta}^{\rho\sigma}+\tilde{\Delta}^{\mu\rho} \tilde{\Delta}^{\nu\sigma}+\tilde{\Delta}^{\mu\sigma} \tilde{\Delta}^{\nu\rho}\right)
 \frac{1}{15}\Bigl[ \mathcal{Z}_0-2\mathcal{Z}_2 + \mathcal{Z}_4 \Bigr]  ,
\end{split}
\end{equation}
where we abbreviate the argument $x$ on $\mathcal{A}^{\mu_1\cdots\mu_n}$ and $\mathcal{Z}_n$.

Let us come back to the evaluation of $\mathcal{K}_1$
With the help of $\mathcal{A}$ and $\mathcal{Z}_0$, we get
\begin{equation}
\begin{split}
 \mathcal{K}_1(y)
  &= -\frac{1}{4\pi^2}\int_0^\infty \frac{\rmd p}{p} \mathcal{A}(py/\hbar)  
  = -\frac{1}{4\pi^2}\int_0^\infty \frac{\rmd p}{p} \mathcal{Z}_0(py/\hbar) \\
  &= -\frac{(-\rmi/\hbar)}{4\pi^2} \int_0^\infty \rmd p
\int_\infty^y \rmd z\, \mathcal{Z}_1(pz/\hbar) \\
  &= -\frac{(-\rmi)}{4\pi^2} 
\int_\infty^y \frac{\rmd z}{z} \cdot \frac{-\rmi}{2} 
  = -\frac{\mathcal{J}}{8\pi^2} 
\end{split}
\end{equation}
with
\begin{equation}
 \mathcal{J} := \int_0^{y^{-1}} \frac{\rmd p}{p}  .
\end{equation}
Here we interchanged the order of integration, as do in the point-splitting regularization for axial anomaly in two dimensions~\cite{Peskin:1995ev}.
The ultraviolet logarithmic divergence is now regularized by the splitting parameter $y$.
The infrared divergence ($z=0$) is to be canceled by the matter part.

In the same manner, we can calculate $\mathcal{K}^{\mu\nu}_2$ and $\mathcal{K}^{\mu\nu\rho\sigma}_3$.
The only extra relation to be utilized is 
\begin{equation}
 \frac{\rmd^n\delta(p^2)}{(\rmd p^2)^n}
 = \frac{(-1)^n\,n!}{\pi} \mathrm{Im} \frac{1}{\,(p^2-\rmi\epsilon)^{n+1}}  .
\end{equation}
Finally, we get the following expressions:
\begin{equation}
\mathcal{K}_2^{\mu\nu}(y) = \frac{\mathcal{J}}{16\pi^2}g^{\mu\nu} ,
\quad
\mathcal{K}_3^{\mu\nu\rho\sigma}(y) 
= -\frac{\mathcal{J}}{32\pi^2} 
(g^{\mu\nu}g^{\rho\sigma}+g^{\mu\rho}g^{\nu\sigma}+g^{\mu\sigma}g^{\nu\rho}) ,
\end{equation}
where we perform the analytic continuation to Minkowski spacetime after integration.

\section{Formulas of background fields}\label{app:Fomega}
In this Appendix, we show several formulas of electromagnetic field and fluid vorticity fields, which are defined in Eq.~\eqref{eq:Fomega}.
The two rank tensors $F_{\mu\nu}$, $\beta^{-1}\partial_\mu \beta_{\nu}$ and their duals are expanded with $E_\mu,B_\mu,a_\mu$ and $\omega_\mu$ as follows:
\begin{equation}
\label{eq:Fomega_expanded}
\begin{split}
 &F_{\mu\nu}
 = E_\mu \xi_\nu - E_\nu \xi_\mu - \varepsilon_{\mu\nu\rho\sigma} B^\rho \xi^\sigma  ,
 \quad
 \tF_{\mu\nu}
 = B_\mu \xi_\nu - B_\nu \xi_\mu + \varepsilon_{\mu\nu\rho\sigma} E^\rho \xi^\sigma  , \\
 &\beta^{-1}\partial_\mu \beta_\nu
 = a_\mu \xi_\nu - a_\nu \xi_\mu - \varepsilon_{\mu\nu\rho\sigma}\omega^\rho \xi^\sigma ,\quad
 \omega_{\mu\nu}
 = \omega_\mu \xi_\nu - \omega_\nu \xi_\mu + \varepsilon_{\mu\nu\rho\sigma}a^\rho\xi^\sigma .
\end{split}
\end{equation}
Thanks to them, one can show $\varepsilon_{\mu\nu\rho\sigma} \omega^\rho E^\sigma = -\varepsilon_{\mu\nu\rho\sigma} a^\rho B^\sigma$ or equivalently,
\begin{equation}
\label{eq:omegaE_aB}
 \omega_{[\mu} E_{\nu]} = -a_{[\mu} B_{\nu]}  .
\end{equation}
Using the equilibrium conditions $\partial_\mu \alpha = F_{\mu\nu}\beta^\nu$ in Eq.~\eqref{eq:equilibrium_f0} and $\beta\cdot \partial F_{\mu\nu} = 0$ in Eq.~\eqref{eq:equilibrium}, we find
\begin{equation}
\label{eq:2nd_partial_mu}
0=\partial_{[\mu}\partial_{\nu]} \alpha = F_{\lambda[\mu}\partial_{\nu]}\beta^\lambda .
\end{equation}
Combined with Eqs.~\eqref{eq:Fomega_expanded} and~\eqref{eq:omegaE_aB}, this leads to
\begin{equation}
 a_{[\mu} E_{\nu]}=\omega_{[\mu}B_{\nu]} = 0  .
\end{equation}
Besides, by using the expanded form~\eqref{eq:Fomega_expanded}, we express the derivatives of the background fields as
\begin{equation}
\begin{split}
\label{eq:partial_EBaomega}
& \partial_\mu E_\nu
 = \xi_\mu\xi_\nu (E\cdot a + B\cdot\omega)
 -g_{\mu\nu}B\cdot\omega + B_\mu \omega_\nu
 + 2\xi_{(\mu}\varepsilon_{\nu)\rho\sigma\lambda}a^\rho B^\sigma \xi^\sigma
 + \xi^\rho \partial_\mu F_{\nu\rho} ,\\
&
 \partial_\mu B_\nu
 =\xi_\mu\xi_\nu (a\cdot B - \omega\cdot E)
 + g_{\mu\nu} \omega\cdot E - E_\mu \omega_\nu
 -2\xi_{(\mu}\varepsilon_{\nu)\rho\sigma\lambda} a^\rho E^\sigma \xi^\lambda
 +\xi^\lambda\partial_\mu \tF_{\nu\lambda} ,\\
& 
 \partial_\mu a_\nu
 = \xi_\mu \xi_\nu (a^2+\omega^2) - g_{\mu\nu} \omega^2 + \omega_\mu\omega_\nu - a_\mu a_\nu 
 + 2\xi_{(\mu}\varepsilon_{\nu)\rho\sigma\lambda}a^\rho\omega^\sigma\xi^\lambda  ,\\
& \partial_\mu \omega_\nu 
 = g_{\mu\nu} a\cdot\omega - 2a_\mu \omega_\nu  .
\end{split}
\end{equation}

\bibliography{2nd_ckt}

\begin{thebibliography}{72}%
\makeatletter
\providecommand \@ifxundefined [1]{%
 \@ifx{#1\undefined}
}%
\providecommand \@ifnum [1]{%
 \ifnum #1\expandafter \@firstoftwo
 \else \expandafter \@secondoftwo
 \fi
}%
\providecommand \@ifx [1]{%
 \ifx #1\expandafter \@firstoftwo
 \else \expandafter \@secondoftwo
 \fi
}%
\providecommand \natexlab [1]{#1}%
\providecommand \enquote  [1]{``#1''}%
\providecommand \bibnamefont  [1]{#1}%
\providecommand \bibfnamefont [1]{#1}%
\providecommand \citenamefont [1]{#1}%
\providecommand \href@noop [0]{\@secondoftwo}%
\providecommand \href [0]{\begingroup \@sanitize@url \@href}%
\providecommand \@href[1]{\@@startlink{#1}\@@href}%
\providecommand \@@href[1]{\endgroup#1\@@endlink}%
\providecommand \@sanitize@url [0]{\catcode `\\12\catcode `\$12\catcode
  `\&12\catcode `\#12\catcode `\^12\catcode `\_12\catcode `\%12\relax}%
\providecommand \@@startlink[1]{}%
\providecommand \@@endlink[0]{}%
\providecommand \url  [0]{\begingroup\@sanitize@url \@url }%
\providecommand \@url [1]{\endgroup\@href {#1}{\urlprefix }}%
\providecommand \urlprefix  [0]{URL }%
\providecommand \Eprint [0]{\href }%
\providecommand \doibase [0]{https://doi.org/}%
\providecommand \selectlanguage [0]{\@gobble}%
\providecommand \bibinfo  [0]{\@secondoftwo}%
\providecommand \bibfield  [0]{\@secondoftwo}%
\providecommand \translation [1]{[#1]}%
\providecommand \BibitemOpen [0]{}%
\providecommand \bibitemStop [0]{}%
\providecommand \bibitemNoStop [0]{.\EOS\space}%
\providecommand \EOS [0]{\spacefactor3000\relax}%
\providecommand \BibitemShut  [1]{\csname bibitem#1\endcsname}%
\let\auto@bib@innerbib\@empty
\bibitem [{\citenamefont {Stephanov}\ and\ \citenamefont
  {Yin}(2012)}]{Stephanov:2012ki}%
  \BibitemOpen
  \bibfield  {author} {\bibinfo {author} {\bibfnamefont {M.~A.}\ \bibnamefont
  {Stephanov}}\ and\ \bibinfo {author} {\bibfnamefont {Y.}~\bibnamefont
  {Yin}},\ }\bibfield  {title} {\bibinfo {title} {{Chiral Kinetic Theory}},\
  }\href {https://doi.org/10.1103/PhysRevLett.109.162001} {\bibfield  {journal}
  {\bibinfo  {journal} {Phys. Rev. Lett.}\ }\textbf {\bibinfo {volume} {109}},\
  \bibinfo {pages} {162001} (\bibinfo {year} {2012})},\ \Eprint
  {https://arxiv.org/abs/1207.0747} {arXiv:1207.0747 [hep-th]} \BibitemShut
  {NoStop}%
\bibitem [{\citenamefont {Son}\ and\ \citenamefont
  {Yamamoto}(2012)}]{Son:2012wh}%
  \BibitemOpen
  \bibfield  {author} {\bibinfo {author} {\bibfnamefont {D.~T.}\ \bibnamefont
  {Son}}\ and\ \bibinfo {author} {\bibfnamefont {N.}~\bibnamefont {Yamamoto}},\
  }\bibfield  {title} {\bibinfo {title} {{Berry Curvature, Triangle Anomalies,
  and the Chiral Magnetic Effect in Fermi Liquids}},\ }\href
  {https://doi.org/10.1103/PhysRevLett.109.181602} {\bibfield  {journal}
  {\bibinfo  {journal} {Phys. Rev. Lett.}\ }\textbf {\bibinfo {volume} {109}},\
  \bibinfo {pages} {181602} (\bibinfo {year} {2012})},\ \Eprint
  {https://arxiv.org/abs/1203.2697} {arXiv:1203.2697 [cond-mat.mes-hall]}
  \BibitemShut {NoStop}%
\bibitem [{\citenamefont {Chen}\ \emph {et~al.}(2013)\citenamefont {Chen},
  \citenamefont {Pu}, \citenamefont {Wang},\ and\ \citenamefont
  {Wang}}]{Chen:2012ca}%
  \BibitemOpen
  \bibfield  {author} {\bibinfo {author} {\bibfnamefont {J.-W.}\ \bibnamefont
  {Chen}}, \bibinfo {author} {\bibfnamefont {S.}~\bibnamefont {Pu}}, \bibinfo
  {author} {\bibfnamefont {Q.}~\bibnamefont {Wang}},\ and\ \bibinfo {author}
  {\bibfnamefont {X.-N.}\ \bibnamefont {Wang}},\ }\bibfield  {title} {\bibinfo
  {title} {{Berry Curvature and Four-Dimensional Monopoles in the Relativistic
  Chiral Kinetic Equation}},\ }\href
  {https://doi.org/10.1103/PhysRevLett.110.262301} {\bibfield  {journal}
  {\bibinfo  {journal} {Phys. Rev. Lett.}\ }\textbf {\bibinfo {volume} {110}},\
  \bibinfo {pages} {262301} (\bibinfo {year} {2013})},\ \Eprint
  {https://arxiv.org/abs/1210.8312} {arXiv:1210.8312 [hep-th]} \BibitemShut
  {NoStop}%
\bibitem [{\citenamefont {Xiao}\ \emph {et~al.}(2010)\citenamefont {Xiao},
  \citenamefont {Chang},\ and\ \citenamefont {Niu}}]{Xiao:2009rm}%
  \BibitemOpen
  \bibfield  {author} {\bibinfo {author} {\bibfnamefont {D.}~\bibnamefont
  {Xiao}}, \bibinfo {author} {\bibfnamefont {M.-C.}\ \bibnamefont {Chang}},\
  and\ \bibinfo {author} {\bibfnamefont {Q.}~\bibnamefont {Niu}},\ }\bibfield
  {title} {\bibinfo {title} {{Berry phase effects on electronic properties}},\
  }\href {https://doi.org/10.1103/RevModPhys.82.1959} {\bibfield  {journal}
  {\bibinfo  {journal} {Rev. Mod. Phys.}\ }\textbf {\bibinfo {volume} {82}},\
  \bibinfo {pages} {1959} (\bibinfo {year} {2010})},\ \Eprint
  {https://arxiv.org/abs/0907.2021} {arXiv:0907.2021 [cond-mat.mes-hall]}
  \BibitemShut {NoStop}%
\bibitem [{\citenamefont {Liu}\ and\ \citenamefont {Yin}(2021)}]{Liu:2021uhn}%
  \BibitemOpen
  \bibfield  {author} {\bibinfo {author} {\bibfnamefont {S.~Y.~F.}\
  \bibnamefont {Liu}}\ and\ \bibinfo {author} {\bibfnamefont {Y.}~\bibnamefont
  {Yin}},\ }\bibfield  {title} {\bibinfo {title} {{Spin polarization induced by
  the hydrodynamic gradients}},\ }\href
  {https://doi.org/10.1007/JHEP07(2021)188} {\bibfield  {journal} {\bibinfo
  {journal} {JHEP}\ }\textbf {\bibinfo {volume} {07}},\ \bibinfo {pages}
  {188}},\ \Eprint {https://arxiv.org/abs/2103.09200} {arXiv:2103.09200
  [hep-ph]} \BibitemShut {NoStop}%
\bibitem [{\citenamefont {Fu}\ \emph {et~al.}(2021)\citenamefont {Fu},
  \citenamefont {Liu}, \citenamefont {Pang}, \citenamefont {Song},\ and\
  \citenamefont {Yin}}]{Fu:2021pok}%
  \BibitemOpen
  \bibfield  {author} {\bibinfo {author} {\bibfnamefont {B.}~\bibnamefont
  {Fu}}, \bibinfo {author} {\bibfnamefont {S.~Y.~F.}\ \bibnamefont {Liu}},
  \bibinfo {author} {\bibfnamefont {L.}~\bibnamefont {Pang}}, \bibinfo {author}
  {\bibfnamefont {H.}~\bibnamefont {Song}},\ and\ \bibinfo {author}
  {\bibfnamefont {Y.}~\bibnamefont {Yin}},\ }\bibfield  {title} {\bibinfo
  {title} {{Shear-Induced Spin Polarization in Heavy-Ion Collisions}},\ }\href
  {https://doi.org/10.1103/PhysRevLett.127.142301} {\bibfield  {journal}
  {\bibinfo  {journal} {Phys. Rev. Lett.}\ }\textbf {\bibinfo {volume} {127}},\
  \bibinfo {pages} {142301} (\bibinfo {year} {2021})},\ \Eprint
  {https://arxiv.org/abs/2103.10403} {arXiv:2103.10403 [hep-ph]} \BibitemShut
  {NoStop}%
\bibitem [{\citenamefont {Gorbar}\ \emph {et~al.}(2016)\citenamefont {Gorbar},
  \citenamefont {Shovkovy}, \citenamefont {Vilchinskii}, \citenamefont
  {Rudenok}, \citenamefont {Boyarsky},\ and\ \citenamefont
  {Ruchayskiy}}]{Gorbar:2016qfh}%
  \BibitemOpen
  \bibfield  {author} {\bibinfo {author} {\bibfnamefont {E.~V.}\ \bibnamefont
  {Gorbar}}, \bibinfo {author} {\bibfnamefont {I.~A.}\ \bibnamefont
  {Shovkovy}}, \bibinfo {author} {\bibfnamefont {S.}~\bibnamefont
  {Vilchinskii}}, \bibinfo {author} {\bibfnamefont {I.}~\bibnamefont
  {Rudenok}}, \bibinfo {author} {\bibfnamefont {A.}~\bibnamefont {Boyarsky}},\
  and\ \bibinfo {author} {\bibfnamefont {O.}~\bibnamefont {Ruchayskiy}},\
  }\bibfield  {title} {\bibinfo {title} {{Anomalous Maxwell equations for
  inhomogeneous chiral plasma}},\ }\href
  {https://doi.org/10.1103/PhysRevD.93.105028} {\bibfield  {journal} {\bibinfo
  {journal} {Phys. Rev. D}\ }\textbf {\bibinfo {volume} {93}},\ \bibinfo
  {pages} {105028} (\bibinfo {year} {2016})},\ \Eprint
  {https://arxiv.org/abs/1603.03442} {arXiv:1603.03442 [hep-th]} \BibitemShut
  {NoStop}%
\bibitem [{\citenamefont {Gorbar}\ \emph
  {et~al.}(2017{\natexlab{a}})\citenamefont {Gorbar}, \citenamefont {Miransky},
  \citenamefont {Shovkovy},\ and\ \citenamefont {Sukhachov}}]{Gorbar:2016ygi}%
  \BibitemOpen
  \bibfield  {author} {\bibinfo {author} {\bibfnamefont {E.~V.}\ \bibnamefont
  {Gorbar}}, \bibinfo {author} {\bibfnamefont {V.~A.}\ \bibnamefont
  {Miransky}}, \bibinfo {author} {\bibfnamefont {I.~A.}\ \bibnamefont
  {Shovkovy}},\ and\ \bibinfo {author} {\bibfnamefont {P.~O.}\ \bibnamefont
  {Sukhachov}},\ }\bibfield  {title} {\bibinfo {title} {{Consistent Chiral
  Kinetic Theory in Weyl Materials: Chiral Magnetic Plasmons}},\ }\href
  {https://doi.org/10.1103/PhysRevLett.118.127601} {\bibfield  {journal}
  {\bibinfo  {journal} {Phys. Rev. Lett.}\ }\textbf {\bibinfo {volume} {118}},\
  \bibinfo {pages} {127601} (\bibinfo {year} {2017}{\natexlab{a}})},\ \Eprint
  {https://arxiv.org/abs/1610.07625} {arXiv:1610.07625 [cond-mat.str-el]}
  \BibitemShut {NoStop}%
\bibitem [{\citenamefont {Yamamoto}(2016)}]{Yamamoto:2015gzz}%
  \BibitemOpen
  \bibfield  {author} {\bibinfo {author} {\bibfnamefont {N.}~\bibnamefont
  {Yamamoto}},\ }\bibfield  {title} {\bibinfo {title} {{Chiral transport of
  neutrinos in supernovae: Neutrino-induced fluid helicity and helical plasma
  instability}},\ }\href {https://doi.org/10.1103/PhysRevD.93.065017}
  {\bibfield  {journal} {\bibinfo  {journal} {Phys. Rev. D}\ }\textbf {\bibinfo
  {volume} {93}},\ \bibinfo {pages} {065017} (\bibinfo {year} {2016})},\
  \Eprint {https://arxiv.org/abs/1511.00933} {arXiv:1511.00933 [astro-ph.HE]}
  \BibitemShut {NoStop}%
\bibitem [{\citenamefont {Yamamoto}\ and\ \citenamefont
  {Yang}(2020)}]{Yamamoto:2020zrs}%
  \BibitemOpen
  \bibfield  {author} {\bibinfo {author} {\bibfnamefont {N.}~\bibnamefont
  {Yamamoto}}\ and\ \bibinfo {author} {\bibfnamefont {D.-L.}\ \bibnamefont
  {Yang}},\ }\bibfield  {title} {\bibinfo {title} {{Chiral Radiation Transport
  Theory of Neutrinos}},\ }\href {https://doi.org/10.3847/1538-4357/ab8468}
  {\bibfield  {journal} {\bibinfo  {journal} {Astrophys. J.}\ }\textbf
  {\bibinfo {volume} {895}},\ \bibinfo {pages} {56} (\bibinfo {year} {2020})},\
  \Eprint {https://arxiv.org/abs/2002.11348} {arXiv:2002.11348 [astro-ph.HE]}
  \BibitemShut {NoStop}%
\bibitem [{\citenamefont {Yamamoto}\ and\ \citenamefont
  {Yang}(2021)}]{Yamamoto:2021hjs}%
  \BibitemOpen
  \bibfield  {author} {\bibinfo {author} {\bibfnamefont {N.}~\bibnamefont
  {Yamamoto}}\ and\ \bibinfo {author} {\bibfnamefont {D.-L.}\ \bibnamefont
  {Yang}},\ }\bibfield  {title} {\bibinfo {title} {{Magnetic field induced
  neutrino chiral transport near equilibrium}},\ }\href
  {https://doi.org/10.1103/PhysRevD.104.123019} {\bibfield  {journal} {\bibinfo
   {journal} {Phys. Rev. D}\ }\textbf {\bibinfo {volume} {104}},\ \bibinfo
  {pages} {123019} (\bibinfo {year} {2021})},\ \Eprint
  {https://arxiv.org/abs/2103.13159} {arXiv:2103.13159 [hep-ph]} \BibitemShut
  {NoStop}%
\bibitem [{\citenamefont {Hattori}\ \emph {et~al.}(2021)\citenamefont
  {Hattori}, \citenamefont {Hidaka}, \citenamefont {Yamamoto},\ and\
  \citenamefont {Yang}}]{Hattori:2020gqh}%
  \BibitemOpen
  \bibfield  {author} {\bibinfo {author} {\bibfnamefont {K.}~\bibnamefont
  {Hattori}}, \bibinfo {author} {\bibfnamefont {Y.}~\bibnamefont {Hidaka}},
  \bibinfo {author} {\bibfnamefont {N.}~\bibnamefont {Yamamoto}},\ and\
  \bibinfo {author} {\bibfnamefont {D.-L.}\ \bibnamefont {Yang}},\ }\bibfield
  {title} {\bibinfo {title} {{Wigner functions and quantum kinetic theory of
  polarized photons}},\ }\href {https://doi.org/10.1007/JHEP02(2021)001}
  {\bibfield  {journal} {\bibinfo  {journal} {JHEP}\ }\textbf {\bibinfo
  {volume} {02}},\ \bibinfo {pages} {001}},\ \Eprint
  {https://arxiv.org/abs/2010.13368} {arXiv:2010.13368 [hep-ph]} \BibitemShut
  {NoStop}%
\bibitem [{\citenamefont {Huang}\ \emph {et~al.}(2020)\citenamefont {Huang},
  \citenamefont {Mitkin}, \citenamefont {Sadofyev},\ and\ \citenamefont
  {Speranza}}]{Huang:2020kik}%
  \BibitemOpen
  \bibfield  {author} {\bibinfo {author} {\bibfnamefont {X.-G.}\ \bibnamefont
  {Huang}}, \bibinfo {author} {\bibfnamefont {P.}~\bibnamefont {Mitkin}},
  \bibinfo {author} {\bibfnamefont {A.~V.}\ \bibnamefont {Sadofyev}},\ and\
  \bibinfo {author} {\bibfnamefont {E.}~\bibnamefont {Speranza}},\ }\bibfield
  {title} {\bibinfo {title} {{Zilch vortical effect, Berry phase, and kinetic
  theory}},\ }\href {https://doi.org/10.1007/JHEP10(2020)117} {\bibfield
  {journal} {\bibinfo  {journal} {JHEP}\ }\textbf {\bibinfo {volume} {10}},\
  \bibinfo {pages} {117}},\ \Eprint {https://arxiv.org/abs/2006.03591}
  {arXiv:2006.03591 [hep-th]} \BibitemShut {NoStop}%
\bibitem [{\citenamefont {Yamamoto}(2017)}]{Yamamoto:2017uul}%
  \BibitemOpen
  \bibfield  {author} {\bibinfo {author} {\bibfnamefont {N.}~\bibnamefont
  {Yamamoto}},\ }\bibfield  {title} {\bibinfo {title} {{Photonic chiral
  vortical effect}},\ }\href {https://doi.org/10.1103/PhysRevD.96.051902}
  {\bibfield  {journal} {\bibinfo  {journal} {Phys. Rev. D}\ }\textbf {\bibinfo
  {volume} {96}},\ \bibinfo {pages} {051902} (\bibinfo {year} {2017})},\
  \Eprint {https://arxiv.org/abs/1702.08886} {arXiv:1702.08886 [hep-th]}
  \BibitemShut {NoStop}%
\bibitem [{\citenamefont {Mameda}\ \emph {et~al.}(2022)\citenamefont {Mameda},
  \citenamefont {Yamamoto},\ and\ \citenamefont {Yang}}]{Mameda:2022ojk}%
  \BibitemOpen
  \bibfield  {author} {\bibinfo {author} {\bibfnamefont {K.}~\bibnamefont
  {Mameda}}, \bibinfo {author} {\bibfnamefont {N.}~\bibnamefont {Yamamoto}},\
  and\ \bibinfo {author} {\bibfnamefont {D.-L.}\ \bibnamefont {Yang}},\
  }\bibfield  {title} {\bibinfo {title} {{Photonic spin Hall effect from
  quantum kinetic theory in curved spacetime}},\ }\href
  {https://doi.org/10.1103/PhysRevD.105.096019} {\bibfield  {journal} {\bibinfo
   {journal} {Phys. Rev. D}\ }\textbf {\bibinfo {volume} {105}},\ \bibinfo
  {pages} {096019} (\bibinfo {year} {2022})},\ \Eprint
  {https://arxiv.org/abs/2203.08449} {arXiv:2203.08449 [hep-th]} \BibitemShut
  {NoStop}%
\bibitem [{\citenamefont {Chen}\ \emph {et~al.}(2014)\citenamefont {Chen},
  \citenamefont {Son}, \citenamefont {Stephanov}, \citenamefont {Yee},\ and\
  \citenamefont {Yin}}]{Chen:2014cla}%
  \BibitemOpen
  \bibfield  {author} {\bibinfo {author} {\bibfnamefont {J.-Y.}\ \bibnamefont
  {Chen}}, \bibinfo {author} {\bibfnamefont {D.~T.}\ \bibnamefont {Son}},
  \bibinfo {author} {\bibfnamefont {M.~A.}\ \bibnamefont {Stephanov}}, \bibinfo
  {author} {\bibfnamefont {H.-U.}\ \bibnamefont {Yee}},\ and\ \bibinfo {author}
  {\bibfnamefont {Y.}~\bibnamefont {Yin}},\ }\bibfield  {title} {\bibinfo
  {title} {{Lorentz Invariance in Chiral Kinetic Theory}},\ }\href
  {https://doi.org/10.1103/PhysRevLett.113.182302} {\bibfield  {journal}
  {\bibinfo  {journal} {Phys. Rev. Lett.}\ }\textbf {\bibinfo {volume} {113}},\
  \bibinfo {pages} {182302} (\bibinfo {year} {2014})},\ \Eprint
  {https://arxiv.org/abs/1404.5963} {arXiv:1404.5963 [hep-th]} \BibitemShut
  {NoStop}%
\bibitem [{\citenamefont {Chen}\ \emph {et~al.}(2015)\citenamefont {Chen},
  \citenamefont {Son},\ and\ \citenamefont {Stephanov}}]{Chen:2015gta}%
  \BibitemOpen
  \bibfield  {author} {\bibinfo {author} {\bibfnamefont {J.-Y.}\ \bibnamefont
  {Chen}}, \bibinfo {author} {\bibfnamefont {D.~T.}\ \bibnamefont {Son}},\ and\
  \bibinfo {author} {\bibfnamefont {M.~A.}\ \bibnamefont {Stephanov}},\
  }\bibfield  {title} {\bibinfo {title} {{Collisions in Chiral Kinetic
  Theory}},\ }\href {https://doi.org/10.1103/PhysRevLett.115.021601} {\bibfield
   {journal} {\bibinfo  {journal} {Phys. Rev. Lett.}\ }\textbf {\bibinfo
  {volume} {115}},\ \bibinfo {pages} {021601} (\bibinfo {year} {2015})},\
  \Eprint {https://arxiv.org/abs/1502.06966} {arXiv:1502.06966 [hep-th]}
  \BibitemShut {NoStop}%
\bibitem [{\citenamefont {Hidaka}\ \emph {et~al.}(2017)\citenamefont {Hidaka},
  \citenamefont {Pu},\ and\ \citenamefont {Yang}}]{Hidaka:2016yjf}%
  \BibitemOpen
  \bibfield  {author} {\bibinfo {author} {\bibfnamefont {Y.}~\bibnamefont
  {Hidaka}}, \bibinfo {author} {\bibfnamefont {S.}~\bibnamefont {Pu}},\ and\
  \bibinfo {author} {\bibfnamefont {D.-L.}\ \bibnamefont {Yang}},\ }\bibfield
  {title} {\bibinfo {title} {{Relativistic chiral kinetic theory from quantum
  field theories}},\ }\href {https://doi.org/10.1103/PhysRevD.95.091901}
  {\bibfield  {journal} {\bibinfo  {journal} {Phys. Rev. D}\ }\textbf {\bibinfo
  {volume} {95}},\ \bibinfo {pages} {091901} (\bibinfo {year} {2017})},\
  \Eprint {https://arxiv.org/abs/1612.04630} {arXiv:1612.04630 [hep-th]}
  \BibitemShut {NoStop}%
\bibitem [{\citenamefont {Yang}\ \emph
  {et~al.}(2020{\natexlab{a}})\citenamefont {Yang}, \citenamefont {Hattori},\
  and\ \citenamefont {Hidaka}}]{Yang:2020hri}%
  \BibitemOpen
  \bibfield  {author} {\bibinfo {author} {\bibfnamefont {D.-L.}\ \bibnamefont
  {Yang}}, \bibinfo {author} {\bibfnamefont {K.}~\bibnamefont {Hattori}},\ and\
  \bibinfo {author} {\bibfnamefont {Y.}~\bibnamefont {Hidaka}},\ }\bibfield
  {title} {\bibinfo {title} {{Effective quantum kinetic theory for spin
  transport of fermions with collsional effects}},\ }\href
  {https://doi.org/10.1007/JHEP07(2020)070} {\bibfield  {journal} {\bibinfo
  {journal} {JHEP}\ }\textbf {\bibinfo {volume} {07}},\ \bibinfo {pages}
  {070}},\ \Eprint {https://arxiv.org/abs/2002.02612} {arXiv:2002.02612
  [hep-ph]} \BibitemShut {NoStop}%
\bibitem [{\citenamefont {Weickgenannt}\ \emph {et~al.}(2021)\citenamefont
  {Weickgenannt}, \citenamefont {Speranza}, \citenamefont {Sheng},
  \citenamefont {Wang},\ and\ \citenamefont {Rischke}}]{Weickgenannt:2021cuo}%
  \BibitemOpen
  \bibfield  {author} {\bibinfo {author} {\bibfnamefont {N.}~\bibnamefont
  {Weickgenannt}}, \bibinfo {author} {\bibfnamefont {E.}~\bibnamefont
  {Speranza}}, \bibinfo {author} {\bibfnamefont {X.-l.}\ \bibnamefont {Sheng}},
  \bibinfo {author} {\bibfnamefont {Q.}~\bibnamefont {Wang}},\ and\ \bibinfo
  {author} {\bibfnamefont {D.~H.}\ \bibnamefont {Rischke}},\ }\bibfield
  {title} {\bibinfo {title} {{Derivation of the nonlocal collision term in the
  relativistic Boltzmann equation for massive spin-1/2 particles from quantum
  field theory}},\ }\href {https://doi.org/10.1103/PhysRevD.104.016022}
  {\bibfield  {journal} {\bibinfo  {journal} {Phys. Rev. D}\ }\textbf {\bibinfo
  {volume} {104}},\ \bibinfo {pages} {016022} (\bibinfo {year} {2021})},\
  \Eprint {https://arxiv.org/abs/2103.04896} {arXiv:2103.04896 [nucl-th]}
  \BibitemShut {NoStop}%
\bibitem [{\citenamefont {Lin}(2022)}]{Lin:2021mvw}%
  \BibitemOpen
  \bibfield  {author} {\bibinfo {author} {\bibfnamefont {S.}~\bibnamefont
  {Lin}},\ }\bibfield  {title} {\bibinfo {title} {{Quantum kinetic theory for
  quantum electrodynamics}},\ }\href
  {https://doi.org/10.1103/PhysRevD.105.076017} {\bibfield  {journal} {\bibinfo
   {journal} {Phys. Rev. D}\ }\textbf {\bibinfo {volume} {105}},\ \bibinfo
  {pages} {076017} (\bibinfo {year} {2022})},\ \Eprint
  {https://arxiv.org/abs/2109.00184} {arXiv:2109.00184 [hep-ph]} \BibitemShut
  {NoStop}%
\bibitem [{\citenamefont {Gao}\ and\ \citenamefont
  {Liang}(2019)}]{Gao:2019znl}%
  \BibitemOpen
  \bibfield  {author} {\bibinfo {author} {\bibfnamefont {J.-H.}\ \bibnamefont
  {Gao}}\ and\ \bibinfo {author} {\bibfnamefont {Z.-T.}\ \bibnamefont
  {Liang}},\ }\bibfield  {title} {\bibinfo {title} {{Relativistic Quantum
  Kinetic Theory for Massive Fermions and Spin Effects}},\ }\href
  {https://doi.org/10.1103/PhysRevD.100.056021} {\bibfield  {journal} {\bibinfo
   {journal} {Phys. Rev. D}\ }\textbf {\bibinfo {volume} {100}},\ \bibinfo
  {pages} {056021} (\bibinfo {year} {2019})},\ \Eprint
  {https://arxiv.org/abs/1902.06510} {arXiv:1902.06510 [hep-ph]} \BibitemShut
  {NoStop}%
\bibitem [{\citenamefont {Weickgenannt}\ \emph {et~al.}(2019)\citenamefont
  {Weickgenannt}, \citenamefont {Sheng}, \citenamefont {Speranza},
  \citenamefont {Wang},\ and\ \citenamefont {Rischke}}]{Weickgenannt:2019dks}%
  \BibitemOpen
  \bibfield  {author} {\bibinfo {author} {\bibfnamefont {N.}~\bibnamefont
  {Weickgenannt}}, \bibinfo {author} {\bibfnamefont {X.-L.}\ \bibnamefont
  {Sheng}}, \bibinfo {author} {\bibfnamefont {E.}~\bibnamefont {Speranza}},
  \bibinfo {author} {\bibfnamefont {Q.}~\bibnamefont {Wang}},\ and\ \bibinfo
  {author} {\bibfnamefont {D.~H.}\ \bibnamefont {Rischke}},\ }\bibfield
  {title} {\bibinfo {title} {{Kinetic theory for massive spin-1/2 particles
  from the Wigner-function formalism}},\ }\href
  {https://doi.org/10.1103/PhysRevD.100.056018} {\bibfield  {journal} {\bibinfo
   {journal} {Phys. Rev. D}\ }\textbf {\bibinfo {volume} {100}},\ \bibinfo
  {pages} {056018} (\bibinfo {year} {2019})},\ \Eprint
  {https://arxiv.org/abs/1902.06513} {arXiv:1902.06513 [hep-ph]} \BibitemShut
  {NoStop}%
\bibitem [{\citenamefont {Hattori}\ \emph {et~al.}(2019)\citenamefont
  {Hattori}, \citenamefont {Hidaka},\ and\ \citenamefont
  {Yang}}]{Hattori:2019ahi}%
  \BibitemOpen
  \bibfield  {author} {\bibinfo {author} {\bibfnamefont {K.}~\bibnamefont
  {Hattori}}, \bibinfo {author} {\bibfnamefont {Y.}~\bibnamefont {Hidaka}},\
  and\ \bibinfo {author} {\bibfnamefont {D.-L.}\ \bibnamefont {Yang}},\
  }\bibfield  {title} {\bibinfo {title} {{Axial Kinetic Theory and Spin
  Transport for Fermions with Arbitrary Mass}},\ }\href
  {https://doi.org/10.1103/PhysRevD.100.096011} {\bibfield  {journal} {\bibinfo
   {journal} {Phys. Rev. D}\ }\textbf {\bibinfo {volume} {100}},\ \bibinfo
  {pages} {096011} (\bibinfo {year} {2019})},\ \Eprint
  {https://arxiv.org/abs/1903.01653} {arXiv:1903.01653 [hep-ph]} \BibitemShut
  {NoStop}%
\bibitem [{\citenamefont {Hattori}\ \emph {et~al.}(2017)\citenamefont
  {Hattori}, \citenamefont {Li}, \citenamefont {Satow},\ and\ \citenamefont
  {Yee}}]{Hattori:2016lqx}%
  \BibitemOpen
  \bibfield  {author} {\bibinfo {author} {\bibfnamefont {K.}~\bibnamefont
  {Hattori}}, \bibinfo {author} {\bibfnamefont {S.}~\bibnamefont {Li}},
  \bibinfo {author} {\bibfnamefont {D.}~\bibnamefont {Satow}},\ and\ \bibinfo
  {author} {\bibfnamefont {H.-U.}\ \bibnamefont {Yee}},\ }\bibfield  {title}
  {\bibinfo {title} {{Longitudinal Conductivity in Strong Magnetic Field in
  Perturbative QCD: Complete Leading Order}},\ }\href
  {https://doi.org/10.1103/PhysRevD.95.076008} {\bibfield  {journal} {\bibinfo
  {journal} {Phys. Rev. D}\ }\textbf {\bibinfo {volume} {95}},\ \bibinfo
  {pages} {076008} (\bibinfo {year} {2017})},\ \Eprint
  {https://arxiv.org/abs/1610.06839} {arXiv:1610.06839 [hep-ph]} \BibitemShut
  {NoStop}%
\bibitem [{\citenamefont {Sheng}\ \emph {et~al.}(2018)\citenamefont {Sheng},
  \citenamefont {Rischke}, \citenamefont {Vasak},\ and\ \citenamefont
  {Wang}}]{Sheng:2017lfu}%
  \BibitemOpen
  \bibfield  {author} {\bibinfo {author} {\bibfnamefont {X.-l.}\ \bibnamefont
  {Sheng}}, \bibinfo {author} {\bibfnamefont {D.~H.}\ \bibnamefont {Rischke}},
  \bibinfo {author} {\bibfnamefont {D.}~\bibnamefont {Vasak}},\ and\ \bibinfo
  {author} {\bibfnamefont {Q.}~\bibnamefont {Wang}},\ }\bibfield  {title}
  {\bibinfo {title} {{Wigner functions for fermions in strong magnetic
  fields}},\ }\href {https://doi.org/10.1140/epja/i2018-12414-9} {\bibfield
  {journal} {\bibinfo  {journal} {Eur. Phys. J. A}\ }\textbf {\bibinfo {volume}
  {54}},\ \bibinfo {pages} {21} (\bibinfo {year} {2018})},\ \Eprint
  {https://arxiv.org/abs/1707.01388} {arXiv:1707.01388 [hep-ph]} \BibitemShut
  {NoStop}%
\bibitem [{\citenamefont {Lin}\ and\ \citenamefont {Yang}(2020)}]{Lin:2019fqo}%
  \BibitemOpen
  \bibfield  {author} {\bibinfo {author} {\bibfnamefont {S.}~\bibnamefont
  {Lin}}\ and\ \bibinfo {author} {\bibfnamefont {L.}~\bibnamefont {Yang}},\
  }\bibfield  {title} {\bibinfo {title} {{Chiral kinetic theory from Landau
  level basis}},\ }\href {https://doi.org/10.1103/PhysRevD.101.034006}
  {\bibfield  {journal} {\bibinfo  {journal} {Phys. Rev. D}\ }\textbf {\bibinfo
  {volume} {101}},\ \bibinfo {pages} {034006} (\bibinfo {year} {2020})},\
  \Eprint {https://arxiv.org/abs/1909.11514} {arXiv:1909.11514 [nucl-th]}
  \BibitemShut {NoStop}%
\bibitem [{\citenamefont {Lin}\ and\ \citenamefont {Yang}(2021)}]{Lin:2021sjw}%
  \BibitemOpen
  \bibfield  {author} {\bibinfo {author} {\bibfnamefont {S.}~\bibnamefont
  {Lin}}\ and\ \bibinfo {author} {\bibfnamefont {L.}~\bibnamefont {Yang}},\
  }\bibfield  {title} {\bibinfo {title} {{Magneto-vortical effect in strong
  magnetic field}},\ }\href {https://doi.org/10.1007/JHEP06(2021)054}
  {\bibfield  {journal} {\bibinfo  {journal} {JHEP}\ }\textbf {\bibinfo
  {volume} {06}},\ \bibinfo {pages} {054}},\ \Eprint
  {https://arxiv.org/abs/2103.11577} {arXiv:2103.11577 [nucl-th]} \BibitemShut
  {NoStop}%
\bibitem [{\citenamefont {Manuel}\ and\ \citenamefont
  {Torres-Rincon}(2014{\natexlab{a}})}]{Manuel:2013zaa}%
  \BibitemOpen
  \bibfield  {author} {\bibinfo {author} {\bibfnamefont {C.}~\bibnamefont
  {Manuel}}\ and\ \bibinfo {author} {\bibfnamefont {J.~M.}\ \bibnamefont
  {Torres-Rincon}},\ }\bibfield  {title} {\bibinfo {title} {{Kinetic theory of
  chiral relativistic plasmas and energy density of their gauge collective
  excitations}},\ }\href {https://doi.org/10.1103/PhysRevD.89.096002}
  {\bibfield  {journal} {\bibinfo  {journal} {Phys. Rev. D}\ }\textbf {\bibinfo
  {volume} {89}},\ \bibinfo {pages} {096002} (\bibinfo {year}
  {2014}{\natexlab{a}})},\ \Eprint {https://arxiv.org/abs/1312.1158}
  {arXiv:1312.1158 [hep-ph]} \BibitemShut {NoStop}%
\bibitem [{\citenamefont {Manuel}\ and\ \citenamefont
  {Torres-Rincon}(2014{\natexlab{b}})}]{Manuel:2014dza}%
  \BibitemOpen
  \bibfield  {author} {\bibinfo {author} {\bibfnamefont {C.}~\bibnamefont
  {Manuel}}\ and\ \bibinfo {author} {\bibfnamefont {J.~M.}\ \bibnamefont
  {Torres-Rincon}},\ }\bibfield  {title} {\bibinfo {title} {{Chiral transport
  equation from the quantum Dirac Hamiltonian and the on-shell effective field
  theory}},\ }\href {https://doi.org/10.1103/PhysRevD.90.076007} {\bibfield
  {journal} {\bibinfo  {journal} {Phys. Rev. D}\ }\textbf {\bibinfo {volume}
  {90}},\ \bibinfo {pages} {076007} (\bibinfo {year} {2014}{\natexlab{b}})},\
  \Eprint {https://arxiv.org/abs/1404.6409} {arXiv:1404.6409 [hep-ph]}
  \BibitemShut {NoStop}%
\bibitem [{\citenamefont {Carignano}\ \emph {et~al.}(2018)\citenamefont
  {Carignano}, \citenamefont {Manuel},\ and\ \citenamefont
  {Torres-Rincon}}]{Carignano:2018gqt}%
  \BibitemOpen
  \bibfield  {author} {\bibinfo {author} {\bibfnamefont {S.}~\bibnamefont
  {Carignano}}, \bibinfo {author} {\bibfnamefont {C.}~\bibnamefont {Manuel}},\
  and\ \bibinfo {author} {\bibfnamefont {J.~M.}\ \bibnamefont
  {Torres-Rincon}},\ }\bibfield  {title} {\bibinfo {title} {{Consistent
  relativistic chiral kinetic theory: A derivation from on-shell effective
  field theory}},\ }\href {https://doi.org/10.1103/PhysRevD.98.076005}
  {\bibfield  {journal} {\bibinfo  {journal} {Phys. Rev. D}\ }\textbf {\bibinfo
  {volume} {98}},\ \bibinfo {pages} {076005} (\bibinfo {year} {2018})},\
  \Eprint {https://arxiv.org/abs/1806.01684} {arXiv:1806.01684 [hep-ph]}
  \BibitemShut {NoStop}%
\bibitem [{\citenamefont {Carignano}\ \emph {et~al.}(2020)\citenamefont
  {Carignano}, \citenamefont {Manuel},\ and\ \citenamefont
  {Torres-Rincon}}]{Carignano:2019zsh}%
  \BibitemOpen
  \bibfield  {author} {\bibinfo {author} {\bibfnamefont {S.}~\bibnamefont
  {Carignano}}, \bibinfo {author} {\bibfnamefont {C.}~\bibnamefont {Manuel}},\
  and\ \bibinfo {author} {\bibfnamefont {J.~M.}\ \bibnamefont
  {Torres-Rincon}},\ }\bibfield  {title} {\bibinfo {title} {{Chiral kinetic
  theory from the on-shell effective field theory: Derivation of collision
  terms}},\ }\href {https://doi.org/10.1103/PhysRevD.102.016003} {\bibfield
  {journal} {\bibinfo  {journal} {Phys. Rev. D}\ }\textbf {\bibinfo {volume}
  {102}},\ \bibinfo {pages} {016003} (\bibinfo {year} {2020})},\ \Eprint
  {https://arxiv.org/abs/1908.00561} {arXiv:1908.00561 [hep-ph]} \BibitemShut
  {NoStop}%
\bibitem [{\citenamefont {Mueller}\ and\ \citenamefont
  {Venugopalan}(2017)}]{Mueller:2017arw}%
  \BibitemOpen
  \bibfield  {author} {\bibinfo {author} {\bibfnamefont {N.}~\bibnamefont
  {Mueller}}\ and\ \bibinfo {author} {\bibfnamefont {R.}~\bibnamefont
  {Venugopalan}},\ }\bibfield  {title} {\bibinfo {title} {{Worldline
  construction of a covariant chiral kinetic theory}},\ }\href
  {https://doi.org/10.1103/PhysRevD.96.016023} {\bibfield  {journal} {\bibinfo
  {journal} {Phys. Rev. D}\ }\textbf {\bibinfo {volume} {96}},\ \bibinfo
  {pages} {016023} (\bibinfo {year} {2017})},\ \Eprint
  {https://arxiv.org/abs/1702.01233} {arXiv:1702.01233 [hep-ph]} \BibitemShut
  {NoStop}%
\bibitem [{\citenamefont {Mueller}\ and\ \citenamefont
  {Venugopalan}(2018)}]{Mueller:2017lzw}%
  \BibitemOpen
  \bibfield  {author} {\bibinfo {author} {\bibfnamefont {N.}~\bibnamefont
  {Mueller}}\ and\ \bibinfo {author} {\bibfnamefont {R.}~\bibnamefont
  {Venugopalan}},\ }\bibfield  {title} {\bibinfo {title} {{The chiral anomaly,
  Berry's phase and chiral kinetic theory, from world-lines in quantum field
  theory}},\ }\href {https://doi.org/10.1103/PhysRevD.97.051901} {\bibfield
  {journal} {\bibinfo  {journal} {Phys. Rev. D}\ }\textbf {\bibinfo {volume}
  {97}},\ \bibinfo {pages} {051901} (\bibinfo {year} {2018})},\ \Eprint
  {https://arxiv.org/abs/1701.03331} {arXiv:1701.03331 [hep-ph]} \BibitemShut
  {NoStop}%
\bibitem [{\citenamefont {Liu}\ \emph {et~al.}(2019)\citenamefont {Liu},
  \citenamefont {Gao}, \citenamefont {Mameda},\ and\ \citenamefont
  {Huang}}]{Liu:2018xip}%
  \BibitemOpen
  \bibfield  {author} {\bibinfo {author} {\bibfnamefont {Y.-C.}\ \bibnamefont
  {Liu}}, \bibinfo {author} {\bibfnamefont {L.-L.}\ \bibnamefont {Gao}},
  \bibinfo {author} {\bibfnamefont {K.}~\bibnamefont {Mameda}},\ and\ \bibinfo
  {author} {\bibfnamefont {X.-G.}\ \bibnamefont {Huang}},\ }\bibfield  {title}
  {\bibinfo {title} {{Chiral kinetic theory in curved spacetime}},\ }\href
  {https://doi.org/10.1103/PhysRevD.99.085014} {\bibfield  {journal} {\bibinfo
  {journal} {Phys. Rev. D}\ }\textbf {\bibinfo {volume} {99}},\ \bibinfo
  {pages} {085014} (\bibinfo {year} {2019})},\ \Eprint
  {https://arxiv.org/abs/1812.10127} {arXiv:1812.10127 [hep-th]} \BibitemShut
  {NoStop}%
\bibitem [{\citenamefont {Liu}\ \emph {et~al.}(2020)\citenamefont {Liu},
  \citenamefont {Mameda},\ and\ \citenamefont {Huang}}]{Liu:2020flb}%
  \BibitemOpen
  \bibfield  {author} {\bibinfo {author} {\bibfnamefont {Y.-C.}\ \bibnamefont
  {Liu}}, \bibinfo {author} {\bibfnamefont {K.}~\bibnamefont {Mameda}},\ and\
  \bibinfo {author} {\bibfnamefont {X.-G.}\ \bibnamefont {Huang}},\ }\bibfield
  {title} {\bibinfo {title} {{Covariant Spin Kinetic Theory I: Collisionless
  Limit}},\ }\href {https://doi.org/10.1088/1674-1137/44/9/094101} {\bibfield
  {journal} {\bibinfo  {journal} {Chin. Phys. C}\ }\textbf {\bibinfo {volume}
  {44}},\ \bibinfo {pages} {094101} (\bibinfo {year} {2020})},\ \Eprint
  {https://arxiv.org/abs/2002.03753} {arXiv:2002.03753 [hep-ph]} \BibitemShut
  {NoStop}%
\bibitem [{\citenamefont {Hayata}\ \emph {et~al.}(2021)\citenamefont {Hayata},
  \citenamefont {Hidaka},\ and\ \citenamefont {Mameda}}]{Hayata:2020sqz}%
  \BibitemOpen
  \bibfield  {author} {\bibinfo {author} {\bibfnamefont {T.}~\bibnamefont
  {Hayata}}, \bibinfo {author} {\bibfnamefont {Y.}~\bibnamefont {Hidaka}},\
  and\ \bibinfo {author} {\bibfnamefont {K.}~\bibnamefont {Mameda}},\
  }\bibfield  {title} {\bibinfo {title} {{Second order chiral kinetic theory
  under gravity and antiparallel charge-energy flow}},\ }\href
  {https://doi.org/10.1007/JHEP05(2021)023} {\bibfield  {journal} {\bibinfo
  {journal} {JHEP}\ }\textbf {\bibinfo {volume} {05}},\ \bibinfo {pages}
  {023}},\ \Eprint {https://arxiv.org/abs/2012.12494} {arXiv:2012.12494
  [hep-th]} \BibitemShut {NoStop}%
\bibitem [{\citenamefont {Gao}\ \emph {et~al.}(2021)\citenamefont {Gao},
  \citenamefont {Kaushik}, \citenamefont {Kharzeev},\ and\ \citenamefont
  {Philip}}]{Gao:2020gcf}%
  \BibitemOpen
  \bibfield  {author} {\bibinfo {author} {\bibfnamefont {L.-L.}\ \bibnamefont
  {Gao}}, \bibinfo {author} {\bibfnamefont {S.}~\bibnamefont {Kaushik}},
  \bibinfo {author} {\bibfnamefont {D.~E.}\ \bibnamefont {Kharzeev}},\ and\
  \bibinfo {author} {\bibfnamefont {E.~J.}\ \bibnamefont {Philip}},\ }\bibfield
   {title} {\bibinfo {title} {{Chiral kinetic theory of anomalous transport
  induced by torsion}},\ }\href {https://doi.org/10.1103/PhysRevB.104.064307}
  {\bibfield  {journal} {\bibinfo  {journal} {Phys. Rev. B}\ }\textbf {\bibinfo
  {volume} {104}},\ \bibinfo {pages} {064307} (\bibinfo {year} {2021})},\
  \Eprint {https://arxiv.org/abs/2010.07123} {arXiv:2010.07123
  [cond-mat.mes-hall]} \BibitemShut {NoStop}%
\bibitem [{\citenamefont {Hidaka}\ \emph {et~al.}(2022)\citenamefont {Hidaka},
  \citenamefont {Pu}, \citenamefont {Wang},\ and\ \citenamefont
  {Yang}}]{Hidaka:2022dmn}%
  \BibitemOpen
  \bibfield  {author} {\bibinfo {author} {\bibfnamefont {Y.}~\bibnamefont
  {Hidaka}}, \bibinfo {author} {\bibfnamefont {S.}~\bibnamefont {Pu}}, \bibinfo
  {author} {\bibfnamefont {Q.}~\bibnamefont {Wang}},\ and\ \bibinfo {author}
  {\bibfnamefont {D.-L.}\ \bibnamefont {Yang}},\ }\bibfield  {title} {\bibinfo
  {title} {{Foundations and applications of quantum kinetic theory}},\ }\href
  {https://doi.org/10.1016/j.ppnp.2022.103989} {\bibfield  {journal} {\bibinfo
  {journal} {Prog. Part. Nucl. Phys.}\ }\textbf {\bibinfo {volume} {127}},\
  \bibinfo {pages} {103989} (\bibinfo {year} {2022})},\ \Eprint
  {https://arxiv.org/abs/2201.07644} {arXiv:2201.07644 [hep-ph]} \BibitemShut
  {NoStop}%
\bibitem [{\citenamefont {Hattori}\ and\ \citenamefont
  {Yin}(2016)}]{Hattori:2016njk}%
  \BibitemOpen
  \bibfield  {author} {\bibinfo {author} {\bibfnamefont {K.}~\bibnamefont
  {Hattori}}\ and\ \bibinfo {author} {\bibfnamefont {Y.}~\bibnamefont {Yin}},\
  }\bibfield  {title} {\bibinfo {title} {{Charge redistribution from anomalous
  magnetovorticity coupling}},\ }\href
  {https://doi.org/10.1103/PhysRevLett.117.152002} {\bibfield  {journal}
  {\bibinfo  {journal} {Phys. Rev. Lett.}\ }\textbf {\bibinfo {volume} {117}},\
  \bibinfo {pages} {152002} (\bibinfo {year} {2016})},\ \Eprint
  {https://arxiv.org/abs/1607.01513} {arXiv:1607.01513 [hep-th]} \BibitemShut
  {NoStop}%
\bibitem [{\citenamefont {Ebihara}\ \emph {et~al.}(2017)\citenamefont
  {Ebihara}, \citenamefont {Fukushima},\ and\ \citenamefont
  {Mameda}}]{Ebihara:2016fwa}%
  \BibitemOpen
  \bibfield  {author} {\bibinfo {author} {\bibfnamefont {S.}~\bibnamefont
  {Ebihara}}, \bibinfo {author} {\bibfnamefont {K.}~\bibnamefont {Fukushima}},\
  and\ \bibinfo {author} {\bibfnamefont {K.}~\bibnamefont {Mameda}},\
  }\bibfield  {title} {\bibinfo {title} {{Boundary effects and gapped
  dispersion in rotating fermionic matter}},\ }\href
  {https://doi.org/10.1016/j.physletb.2016.11.010} {\bibfield  {journal}
  {\bibinfo  {journal} {Phys. Lett. B}\ }\textbf {\bibinfo {volume} {764}},\
  \bibinfo {pages} {94} (\bibinfo {year} {2017})},\ \Eprint
  {https://arxiv.org/abs/1608.00336} {arXiv:1608.00336 [hep-ph]} \BibitemShut
  {NoStop}%
\bibitem [{\citenamefont {Kharzeev}\ and\ \citenamefont
  {Yee}(2011)}]{Kharzeev:2011ds}%
  \BibitemOpen
  \bibfield  {author} {\bibinfo {author} {\bibfnamefont {D.~E.}\ \bibnamefont
  {Kharzeev}}\ and\ \bibinfo {author} {\bibfnamefont {H.-U.}\ \bibnamefont
  {Yee}},\ }\bibfield  {title} {\bibinfo {title} {{Anomalies and time reversal
  invariance in relativistic hydrodynamics: the second order and higher
  dimensional formulations}},\ }\href
  {https://doi.org/10.1103/PhysRevD.84.045025} {\bibfield  {journal} {\bibinfo
  {journal} {Phys. Rev. D}\ }\textbf {\bibinfo {volume} {84}},\ \bibinfo
  {pages} {045025} (\bibinfo {year} {2011})},\ \Eprint
  {https://arxiv.org/abs/1105.6360} {arXiv:1105.6360 [hep-th]} \BibitemShut
  {NoStop}%
\bibitem [{\citenamefont {Elze}\ \emph {et~al.}(1986)\citenamefont {Elze},
  \citenamefont {Gyulassy},\ and\ \citenamefont {Vasak}}]{Elze:1986qd}%
  \BibitemOpen
  \bibfield  {author} {\bibinfo {author} {\bibfnamefont {H.~T.}\ \bibnamefont
  {Elze}}, \bibinfo {author} {\bibfnamefont {M.}~\bibnamefont {Gyulassy}},\
  and\ \bibinfo {author} {\bibfnamefont {D.}~\bibnamefont {Vasak}},\ }\bibfield
   {title} {\bibinfo {title} {{Transport Equations for the {QCD} Quark Wigner
  Operator}},\ }\href {https://doi.org/10.1016/0550-3213(86)90072-6} {\bibfield
   {journal} {\bibinfo  {journal} {Nucl. Phys.}\ }\textbf {\bibinfo {volume}
  {B276}},\ \bibinfo {pages} {706} (\bibinfo {year} {1986})}\BibitemShut
  {NoStop}%
\bibitem [{\citenamefont {Yang}\ \emph
  {et~al.}(2020{\natexlab{b}})\citenamefont {Yang}, \citenamefont {Gao},
  \citenamefont {Liang},\ and\ \citenamefont {Wang}}]{Yang:2020mtz}%
  \BibitemOpen
  \bibfield  {author} {\bibinfo {author} {\bibfnamefont {S.-Z.}\ \bibnamefont
  {Yang}}, \bibinfo {author} {\bibfnamefont {J.-H.}\ \bibnamefont {Gao}},
  \bibinfo {author} {\bibfnamefont {Z.-T.}\ \bibnamefont {Liang}},\ and\
  \bibinfo {author} {\bibfnamefont {Q.}~\bibnamefont {Wang}},\ }\bibfield
  {title} {\bibinfo {title} {{Second-order charge currents and stress tensor in
  a chiral system}},\ }\href {https://doi.org/10.1103/PhysRevD.102.116024}
  {\bibfield  {journal} {\bibinfo  {journal} {Phys. Rev. D}\ }\textbf {\bibinfo
  {volume} {102}},\ \bibinfo {pages} {116024} (\bibinfo {year}
  {2020}{\natexlab{b}})},\ \Eprint {https://arxiv.org/abs/2003.04517}
  {arXiv:2003.04517 [hep-ph]} \BibitemShut {NoStop}%
\bibitem [{\citenamefont {Heisenberg}\ and\ \citenamefont
  {Euler}(1936)}]{Heisenberg:1936nmg}%
  \BibitemOpen
  \bibfield  {author} {\bibinfo {author} {\bibfnamefont {W.}~\bibnamefont
  {Heisenberg}}\ and\ \bibinfo {author} {\bibfnamefont {H.}~\bibnamefont
  {Euler}},\ }\bibfield  {title} {\bibinfo {title} {{Consequences of Dirac's
  theory of positrons}},\ }\href {https://doi.org/10.1007/BF01343663}
  {\bibfield  {journal} {\bibinfo  {journal} {Z. Phys.}\ }\textbf {\bibinfo
  {volume} {98}},\ \bibinfo {pages} {714} (\bibinfo {year} {1936})},\ \Eprint
  {https://arxiv.org/abs/physics/0605038} {arXiv:physics/0605038} \BibitemShut
  {NoStop}%
\bibitem [{\citenamefont {Schwinger}(1951)}]{PhysRev.82.664}%
  \BibitemOpen
  \bibfield  {author} {\bibinfo {author} {\bibfnamefont {J.}~\bibnamefont
  {Schwinger}},\ }\bibfield  {title} {\bibinfo {title} {{On Gauge Invariance
  and Vacuum Polarization}},\ }\href {https://doi.org/10.1103/PhysRev.82.664}
  {\bibfield  {journal} {\bibinfo  {journal} {Phys. Rev.}\ }\textbf {\bibinfo
  {volume} {82}},\ \bibinfo {pages} {664} (\bibinfo {year} {1951})}\BibitemShut
  {NoStop}%
\bibitem [{\citenamefont {Schwinger}(1962)}]{PhysRev.128.2425}%
  \BibitemOpen
  \bibfield  {author} {\bibinfo {author} {\bibfnamefont {J.}~\bibnamefont
  {Schwinger}},\ }\bibfield  {title} {\bibinfo {title} {{Gauge Invariance and
  Mass. II}},\ }\href {https://doi.org/10.1103/PhysRev.128.2425} {\bibfield
  {journal} {\bibinfo  {journal} {Phys. Rev.}\ }\textbf {\bibinfo {volume}
  {128}},\ \bibinfo {pages} {2425} (\bibinfo {year} {1962})}\BibitemShut
  {NoStop}%
\bibitem [{\citenamefont {Bellac}(2011)}]{Bellac:2011kqa}%
  \BibitemOpen
  \bibfield  {author} {\bibinfo {author} {\bibfnamefont {M.~L.}\ \bibnamefont
  {Bellac}},\ }\href@noop {} {\emph {\bibinfo {title} {{Thermal Field
  Theory}}}}\ (\bibinfo  {publisher} {Cambridge University Press},\ \bibinfo
  {year} {2011})\BibitemShut {NoStop}%
\bibitem [{\citenamefont {'t~Hooft}\ and\ \citenamefont
  {Veltman}(1972)}]{tHooft:1972tcz}%
  \BibitemOpen
  \bibfield  {author} {\bibinfo {author} {\bibfnamefont {G.}~\bibnamefont
  {'t~Hooft}}\ and\ \bibinfo {author} {\bibfnamefont {M.~J.~G.}\ \bibnamefont
  {Veltman}},\ }\bibfield  {title} {\bibinfo {title} {{Regularization and
  Renormalization of Gauge Fields}},\ }\href
  {https://doi.org/10.1016/0550-3213(72)90279-9} {\bibfield  {journal}
  {\bibinfo  {journal} {Nucl. Phys. B}\ }\textbf {\bibinfo {volume} {44}},\
  \bibinfo {pages} {189} (\bibinfo {year} {1972})}\BibitemShut {NoStop}%
\bibitem [{\citenamefont {Peskin}\ and\ \citenamefont
  {Schroeder}(1995)}]{Peskin:1995ev}%
  \BibitemOpen
  \bibfield  {author} {\bibinfo {author} {\bibfnamefont {M.~E.}\ \bibnamefont
  {Peskin}}\ and\ \bibinfo {author} {\bibfnamefont {D.~V.}\ \bibnamefont
  {Schroeder}},\ }\href@noop {} {\emph {\bibinfo {title} {{An Introduction to
  quantum field theory}}}}\ (\bibinfo  {publisher} {Addison-Wesley},\ \bibinfo
  {year} {1995})\BibitemShut {NoStop}%
\bibitem [{\citenamefont {Vilenkin}(1980)}]{Vilenkin:1980fu}%
  \BibitemOpen
  \bibfield  {author} {\bibinfo {author} {\bibfnamefont {A.}~\bibnamefont
  {Vilenkin}},\ }\bibfield  {title} {\bibinfo {title} {{Equilibrium parity
  violating current in a magnetic field}},\ }\href
  {https://doi.org/10.1103/PhysRevD.22.3080} {\bibfield  {journal} {\bibinfo
  {journal} {Phys. Rev. D}\ }\textbf {\bibinfo {volume} {22}},\ \bibinfo
  {pages} {3080} (\bibinfo {year} {1980})}\BibitemShut {NoStop}%
\bibitem [{\citenamefont {Nielsen}\ and\ \citenamefont
  {Ninomiya}(1983)}]{Nielsen:1983rb}%
  \BibitemOpen
  \bibfield  {author} {\bibinfo {author} {\bibfnamefont {H.~B.}\ \bibnamefont
  {Nielsen}}\ and\ \bibinfo {author} {\bibfnamefont {M.}~\bibnamefont
  {Ninomiya}},\ }\bibfield  {title} {\bibinfo {title} {{Adler-Bell-Jackiw
  anomaly and Weyl fermions in crystal}},\ }\href
  {https://doi.org/10.1016/0370-2693(83)91529-0} {\bibfield  {journal}
  {\bibinfo  {journal} {Phys. Lett. B}\ }\textbf {\bibinfo {volume} {130}},\
  \bibinfo {pages} {389} (\bibinfo {year} {1983})}\BibitemShut {NoStop}%
\bibitem [{\citenamefont {Fukushima}\ \emph {et~al.}(2008)\citenamefont
  {Fukushima}, \citenamefont {Kharzeev},\ and\ \citenamefont
  {Warringa}}]{Fukushima:2008xe}%
  \BibitemOpen
  \bibfield  {author} {\bibinfo {author} {\bibfnamefont {K.}~\bibnamefont
  {Fukushima}}, \bibinfo {author} {\bibfnamefont {D.~E.}\ \bibnamefont
  {Kharzeev}},\ and\ \bibinfo {author} {\bibfnamefont {H.~J.}\ \bibnamefont
  {Warringa}},\ }\bibfield  {title} {\bibinfo {title} {{The chiral magnetic
  effect}},\ }\href {https://doi.org/10.1103/PhysRevD.78.074033} {\bibfield
  {journal} {\bibinfo  {journal} {Phys. Rev. D}\ }\textbf {\bibinfo {volume}
  {78}},\ \bibinfo {pages} {074033} (\bibinfo {year} {2008})},\ \Eprint
  {https://arxiv.org/abs/0808.3382} {arXiv:0808.3382 [hep-ph]} \BibitemShut
  {NoStop}%
\bibitem [{\citenamefont {Vilenkin}(1979)}]{Vilenkin:1979ui}%
  \BibitemOpen
  \bibfield  {author} {\bibinfo {author} {\bibfnamefont {A.}~\bibnamefont
  {Vilenkin}},\ }\bibfield  {title} {\bibinfo {title} {{Macroscopic parity
  violating effects: Neutrino fluxes from rotating black holes and in rotating
  thermal radiation}},\ }\href {https://doi.org/10.1103/PhysRevD.20.1807}
  {\bibfield  {journal} {\bibinfo  {journal} {Phys. Rev. D}\ }\textbf {\bibinfo
  {volume} {20}},\ \bibinfo {pages} {1807} (\bibinfo {year}
  {1979})}\BibitemShut {NoStop}%
\bibitem [{\citenamefont {Son}\ and\ \citenamefont
  {Sur\'owka}(2009)}]{Son:2009tf}%
  \BibitemOpen
  \bibfield  {author} {\bibinfo {author} {\bibfnamefont {D.~T.}\ \bibnamefont
  {Son}}\ and\ \bibinfo {author} {\bibfnamefont {P.}~\bibnamefont
  {Sur\'owka}},\ }\bibfield  {title} {\bibinfo {title} {{Hydrodynamics with
  Triangle Anomalies}},\ }\href
  {https://doi.org/10.1103/PhysRevLett.103.191601} {\bibfield  {journal}
  {\bibinfo  {journal} {Phys. Rev. Lett.}\ }\textbf {\bibinfo {volume} {103}},\
  \bibinfo {pages} {191601} (\bibinfo {year} {2009})},\ \Eprint
  {https://arxiv.org/abs/0906.5044} {arXiv:0906.5044 [hep-th]} \BibitemShut
  {NoStop}%
\bibitem [{\citenamefont {Landsteiner}\ \emph {et~al.}(2011)\citenamefont
  {Landsteiner}, \citenamefont {Megias},\ and\ \citenamefont
  {Pena-Benitez}}]{Landsteiner:2011cp}%
  \BibitemOpen
  \bibfield  {author} {\bibinfo {author} {\bibfnamefont {K.}~\bibnamefont
  {Landsteiner}}, \bibinfo {author} {\bibfnamefont {E.}~\bibnamefont
  {Megias}},\ and\ \bibinfo {author} {\bibfnamefont {F.}~\bibnamefont
  {Pena-Benitez}},\ }\bibfield  {title} {\bibinfo {title} {{Gravitational
  Anomaly and Transport}},\ }\href
  {https://doi.org/10.1103/PhysRevLett.107.021601} {\bibfield  {journal}
  {\bibinfo  {journal} {Phys. Rev. Lett.}\ }\textbf {\bibinfo {volume} {107}},\
  \bibinfo {pages} {021601} (\bibinfo {year} {2011})},\ \Eprint
  {https://arxiv.org/abs/1103.5006} {arXiv:1103.5006 [hep-ph]} \BibitemShut
  {NoStop}%
\bibitem [{\citenamefont {Bu}\ and\ \citenamefont {Lin}(2020)}]{Bu:2019qmd}%
  \BibitemOpen
  \bibfield  {author} {\bibinfo {author} {\bibfnamefont {Y.}~\bibnamefont
  {Bu}}\ and\ \bibinfo {author} {\bibfnamefont {S.}~\bibnamefont {Lin}},\
  }\bibfield  {title} {\bibinfo {title} {{Magneto-vortical effect in strongly
  coupled plasma}},\ }\href {https://doi.org/10.1140/epjc/s10052-020-7951-5}
  {\bibfield  {journal} {\bibinfo  {journal} {Eur. Phys. J. C}\ }\textbf
  {\bibinfo {volume} {80}},\ \bibinfo {pages} {401} (\bibinfo {year} {2020})},\
  \Eprint {https://arxiv.org/abs/1912.11277} {arXiv:1912.11277 [hep-th]}
  \BibitemShut {NoStop}%
\bibitem [{\citenamefont {Giannotti}\ and\ \citenamefont
  {Mottola}(2009)}]{Giannotti:2008cv}%
  \BibitemOpen
  \bibfield  {author} {\bibinfo {author} {\bibfnamefont {M.}~\bibnamefont
  {Giannotti}}\ and\ \bibinfo {author} {\bibfnamefont {E.}~\bibnamefont
  {Mottola}},\ }\bibfield  {title} {\bibinfo {title} {{The Trace Anomaly and
  Massless Scalar Degrees of Freedom in Gravity}},\ }\href
  {https://doi.org/10.1103/PhysRevD.79.045014} {\bibfield  {journal} {\bibinfo
  {journal} {Phys. Rev. D}\ }\textbf {\bibinfo {volume} {79}},\ \bibinfo
  {pages} {045014} (\bibinfo {year} {2009})},\ \Eprint
  {https://arxiv.org/abs/0812.0351} {arXiv:0812.0351 [hep-th]} \BibitemShut
  {NoStop}%
\bibitem [{\citenamefont {Bastianelli}\ and\ \citenamefont
  {Broccoli}(2019)}]{Bastianelli:2018osv}%
  \BibitemOpen
  \bibfield  {author} {\bibinfo {author} {\bibfnamefont {F.}~\bibnamefont
  {Bastianelli}}\ and\ \bibinfo {author} {\bibfnamefont {M.}~\bibnamefont
  {Broccoli}},\ }\bibfield  {title} {\bibinfo {title} {{On the trace anomaly of
  a Weyl fermion in a gauge background}},\ }\href
  {https://doi.org/10.1140/epjc/s10052-019-6799-z} {\bibfield  {journal}
  {\bibinfo  {journal} {Eur. Phys. J. C}\ }\textbf {\bibinfo {volume} {79}},\
  \bibinfo {pages} {292} (\bibinfo {year} {2019})},\ \Eprint
  {https://arxiv.org/abs/1808.03489} {arXiv:1808.03489 [hep-th]} \BibitemShut
  {NoStop}%
\bibitem [{\citenamefont {Bastianelli}\ and\ \citenamefont
  {Chiese}(2022)}]{Bastianelli:2022hmu}%
  \BibitemOpen
  \bibfield  {author} {\bibinfo {author} {\bibfnamefont {F.}~\bibnamefont
  {Bastianelli}}\ and\ \bibinfo {author} {\bibfnamefont {L.}~\bibnamefont
  {Chiese}},\ }\bibfield  {title} {\bibinfo {title} {{Chiral fermions,
  dimensional regularization, and the trace anomaly}},\ }\href
  {https://doi.org/10.1016/j.nuclphysb.2022.115914} {\bibfield  {journal}
  {\bibinfo  {journal} {Nucl. Phys. B}\ }\textbf {\bibinfo {volume} {983}},\
  \bibinfo {pages} {115914} (\bibinfo {year} {2022})},\ \Eprint
  {https://arxiv.org/abs/2203.11668} {arXiv:2203.11668 [hep-th]} \BibitemShut
  {NoStop}%
\bibitem [{\citenamefont {Lee}\ \emph {et~al.}(1989)\citenamefont {Lee},
  \citenamefont {Pac},\ and\ \citenamefont {Shin}}]{Lee:1989vh}%
  \BibitemOpen
  \bibfield  {author} {\bibinfo {author} {\bibfnamefont {H.~W.}\ \bibnamefont
  {Lee}}, \bibinfo {author} {\bibfnamefont {P.~Y.}\ \bibnamefont {Pac}},\ and\
  \bibinfo {author} {\bibfnamefont {H.~K.}\ \bibnamefont {Shin}},\ }\bibfield
  {title} {\bibinfo {title} {{Derivative expansions in quantum
  electrodynamics}},\ }\href {https://doi.org/10.1103/PhysRevD.40.4202}
  {\bibfield  {journal} {\bibinfo  {journal} {Phys. Rev. D}\ }\textbf {\bibinfo
  {volume} {40}},\ \bibinfo {pages} {4202} (\bibinfo {year}
  {1989})}\BibitemShut {NoStop}%
\bibitem [{\citenamefont {Gusynin}\ and\ \citenamefont
  {Shovkovy}(1996)}]{Gusynin:1995bc}%
  \BibitemOpen
  \bibfield  {author} {\bibinfo {author} {\bibfnamefont {V.~P.}\ \bibnamefont
  {Gusynin}}\ and\ \bibinfo {author} {\bibfnamefont {I.~A.}\ \bibnamefont
  {Shovkovy}},\ }\bibfield  {title} {\bibinfo {title} {{Derivative expansion
  for the one loop effective Lagrangian in QED}},\ }\href
  {https://doi.org/10.1139/p96-044} {\bibfield  {journal} {\bibinfo  {journal}
  {Can. J. Phys.}\ }\textbf {\bibinfo {volume} {74}},\ \bibinfo {pages} {282}
  (\bibinfo {year} {1996})},\ \Eprint {https://arxiv.org/abs/hep-ph/9509383}
  {arXiv:hep-ph/9509383} \BibitemShut {NoStop}%
\bibitem [{\citenamefont {Schwartz}(2014)}]{Schwartz:2014sze}%
  \BibitemOpen
  \bibfield  {author} {\bibinfo {author} {\bibfnamefont {M.~D.}\ \bibnamefont
  {Schwartz}},\ }\href@noop {} {\emph {\bibinfo {title} {{Quantum Field Theory
  and the Standard Model}}}}\ (\bibinfo  {publisher} {Cambridge University
  Press},\ \bibinfo {year} {2014})\BibitemShut {NoStop}%
\bibitem [{\citenamefont {Boyd}(2020)}]{boyd2020nonlinear}%
  \BibitemOpen
  \bibfield  {author} {\bibinfo {author} {\bibfnamefont {R.~W.}\ \bibnamefont
  {Boyd}},\ }\href@noop {} {\emph {\bibinfo {title} {Nonlinear optics}}}\
  (\bibinfo  {publisher} {Academic press},\ \bibinfo {year} {2020})\BibitemShut
  {NoStop}%
\bibitem [{\citenamefont {Sodemann}\ and\ \citenamefont
  {Fu}(2015)}]{PhysRevLett.115.216806}%
  \BibitemOpen
  \bibfield  {author} {\bibinfo {author} {\bibfnamefont {I.}~\bibnamefont
  {Sodemann}}\ and\ \bibinfo {author} {\bibfnamefont {L.}~\bibnamefont {Fu}},\
  }\bibfield  {title} {\bibinfo {title} {Quantum nonlinear hall effect induced
  by berry curvature dipole in time-reversal invariant materials},\ }\href
  {https://doi.org/10.1103/PhysRevLett.115.216806} {\bibfield  {journal}
  {\bibinfo  {journal} {Phys. Rev. Lett.}\ }\textbf {\bibinfo {volume} {115}},\
  \bibinfo {pages} {216806} (\bibinfo {year} {2015})}\BibitemShut {NoStop}%
\bibitem [{\citenamefont {Du}\ \emph {et~al.}(2021)\citenamefont {Du},
  \citenamefont {Lu},\ and\ \citenamefont {Xie}}]{du2021nonlinear}%
  \BibitemOpen
  \bibfield  {author} {\bibinfo {author} {\bibfnamefont {Z.}~\bibnamefont
  {Du}}, \bibinfo {author} {\bibfnamefont {H.-Z.}\ \bibnamefont {Lu}},\ and\
  \bibinfo {author} {\bibfnamefont {X.}~\bibnamefont {Xie}},\ }\bibfield
  {title} {\bibinfo {title} {Nonlinear hall effects},\ }\href@noop {}
  {\bibfield  {journal} {\bibinfo  {journal} {Nature Reviews Physics}\ }\textbf
  {\bibinfo {volume} {3}},\ \bibinfo {pages} {744} (\bibinfo {year}
  {2021})}\BibitemShut {NoStop}%
\bibitem [{\citenamefont {Gao}\ \emph {et~al.}(2014)\citenamefont {Gao},
  \citenamefont {Yang},\ and\ \citenamefont {Niu}}]{PhysRevLett.112.166601}%
  \BibitemOpen
  \bibfield  {author} {\bibinfo {author} {\bibfnamefont {Y.}~\bibnamefont
  {Gao}}, \bibinfo {author} {\bibfnamefont {S.~A.}\ \bibnamefont {Yang}},\ and\
  \bibinfo {author} {\bibfnamefont {Q.}~\bibnamefont {Niu}},\ }\bibfield
  {title} {\bibinfo {title} {Field induced positional shift of bloch electrons
  and its dynamical implications},\ }\href
  {https://doi.org/10.1103/PhysRevLett.112.166601} {\bibfield  {journal}
  {\bibinfo  {journal} {Phys. Rev. Lett.}\ }\textbf {\bibinfo {volume} {112}},\
  \bibinfo {pages} {166601} (\bibinfo {year} {2014})}\BibitemShut {NoStop}%
\bibitem [{\citenamefont {Gao}\ \emph {et~al.}(2015)\citenamefont {Gao},
  \citenamefont {Yang},\ and\ \citenamefont {Niu}}]{PhysRevB.91.214405}%
  \BibitemOpen
  \bibfield  {author} {\bibinfo {author} {\bibfnamefont {Y.}~\bibnamefont
  {Gao}}, \bibinfo {author} {\bibfnamefont {S.~A.}\ \bibnamefont {Yang}},\ and\
  \bibinfo {author} {\bibfnamefont {Q.}~\bibnamefont {Niu}},\ }\bibfield
  {title} {\bibinfo {title} {Geometrical effects in orbital magnetic
  susceptibility},\ }\href {https://doi.org/10.1103/PhysRevB.91.214405}
  {\bibfield  {journal} {\bibinfo  {journal} {Phys. Rev. B}\ }\textbf {\bibinfo
  {volume} {91}},\ \bibinfo {pages} {214405} (\bibinfo {year}
  {2015})}\BibitemShut {NoStop}%
\bibitem [{\citenamefont {Gorbar}\ \emph
  {et~al.}(2017{\natexlab{b}})\citenamefont {Gorbar}, \citenamefont {Miransky},
  \citenamefont {Shovkovy},\ and\ \citenamefont {Sukhachov}}]{Gorbar:2017cwv}%
  \BibitemOpen
  \bibfield  {author} {\bibinfo {author} {\bibfnamefont {E.~V.}\ \bibnamefont
  {Gorbar}}, \bibinfo {author} {\bibfnamefont {V.~A.}\ \bibnamefont
  {Miransky}}, \bibinfo {author} {\bibfnamefont {I.~A.}\ \bibnamefont
  {Shovkovy}},\ and\ \bibinfo {author} {\bibfnamefont {P.~O.}\ \bibnamefont
  {Sukhachov}},\ }\bibfield  {title} {\bibinfo {title} {{Second-order chiral
  kinetic theory: Chiral magnetic and pseudomagnetic waves}},\ }\href
  {https://doi.org/10.1103/PhysRevB.95.205141} {\bibfield  {journal} {\bibinfo
  {journal} {Phys. Rev. B}\ }\textbf {\bibinfo {volume} {95}},\ \bibinfo
  {pages} {205141} (\bibinfo {year} {2017}{\natexlab{b}})},\ \Eprint
  {https://arxiv.org/abs/1702.02950} {arXiv:1702.02950 [cond-mat.mes-hall]}
  \BibitemShut {NoStop}%
\bibitem [{\citenamefont {Kadanoff}\ and\ \citenamefont
  {Baym}(1962)}]{kadanoff1962quantum}%
  \BibitemOpen
  \bibfield  {author} {\bibinfo {author} {\bibfnamefont {L.}~\bibnamefont
  {Kadanoff}}\ and\ \bibinfo {author} {\bibfnamefont {G.}~\bibnamefont
  {Baym}},\ }\href@noop {} {\emph {\bibinfo {title} {Quantum Statistical
  Mechanics: Green's Function Methods in Equilibrium and Nonequilibrium
  Problems}}}\ (\bibinfo  {publisher} {Benjamin},\ \bibinfo {year}
  {1962})\BibitemShut {NoStop}%
\bibitem [{\citenamefont {Blaizot}\ and\ \citenamefont
  {Iancu}(2002)}]{Blaizot:2001nr}%
  \BibitemOpen
  \bibfield  {author} {\bibinfo {author} {\bibfnamefont {J.-P.}\ \bibnamefont
  {Blaizot}}\ and\ \bibinfo {author} {\bibfnamefont {E.}~\bibnamefont
  {Iancu}},\ }\bibfield  {title} {\bibinfo {title} {{The Quark gluon plasma:
  Collective dynamics and hard thermal loops}},\ }\href
  {https://doi.org/10.1016/S0370-1573(01)00061-8} {\bibfield  {journal}
  {\bibinfo  {journal} {Phys. Rept.}\ }\textbf {\bibinfo {volume} {359}},\
  \bibinfo {pages} {355} (\bibinfo {year} {2002})},\ \Eprint
  {https://arxiv.org/abs/hep-ph/0101103} {arXiv:hep-ph/0101103} \BibitemShut
  {NoStop}%
\bibitem [{\citenamefont {Akamatsu}\ and\ \citenamefont
  {Yamamoto}(2013)}]{Akamatsu:2013pjd}%
  \BibitemOpen
  \bibfield  {author} {\bibinfo {author} {\bibfnamefont {Y.}~\bibnamefont
  {Akamatsu}}\ and\ \bibinfo {author} {\bibfnamefont {N.}~\bibnamefont
  {Yamamoto}},\ }\bibfield  {title} {\bibinfo {title} {{Chiral Plasma
  Instabilities}},\ }\href {https://doi.org/10.1103/PhysRevLett.111.052002}
  {\bibfield  {journal} {\bibinfo  {journal} {Phys. Rev. Lett.}\ }\textbf
  {\bibinfo {volume} {111}},\ \bibinfo {pages} {052002} (\bibinfo {year}
  {2013})},\ \Eprint {https://arxiv.org/abs/1302.2125} {arXiv:1302.2125
  [nucl-th]} \BibitemShut {NoStop}%
\end{thebibliography}%

\end{document}